\documentclass[iop]{emulateapj}  

\usepackage{color}
\usepackage{hyperref}  
\usepackage{breakurl}  

\newcommand\kms{\ifmmode{\rm km\thinspace s^{-1}}\else km\thinspace s$^{-1}$\fi}
\newcommand\epic{EPIC~219552514}
\newcommand\epicsecond{EPIC~219568666}
\newcommand\epicfirst{EPIC~219394517}
\newcommand\rup{Ruprecht~147}
\newcommand\ktwo{{\it K2\/}}
\newcommand\kepler{{\it Kepler\/}}

\shortauthors{Torres et al.}
\shorttitle{Eclipsing binary in \rup}

\begin{document} 
\submitted{Accepted for publication in The Astrophysical Journal}

\title{Eclipsing binaries in the open cluster Ruprecht 147. III: The
  triple system \epic\ at the main-sequence turnoff}

\author{
Guillermo Torres\altaffilmark{1},
Andrew Vanderburg\altaffilmark{2,3},
Jason L.\ Curtis\altaffilmark{4},
Adam L.\ Kraus\altaffilmark{2},
Aaron C.\ Rizzuto\altaffilmark{2}, and
Michael J. Ireland\altaffilmark{5}
}

\altaffiltext{1}{Center for Astrophysics $\vert$ Harvard \&
  Smithsonian, 60 Garden St., Cambridge, MA 02138, USA;
  gtorres@cfa.harvard.edu}

\altaffiltext{2}{Department of Astronomy, The University of Texas at
  Austin, Austin, TX 78712, USA}

\altaffiltext{3}{NASA Sagan Fellow}

\altaffiltext{4}{American Museum of Natural History, Central Park
  West, New York, NY, USA}

\altaffiltext{5}{Research School of Astronomy and Astrophysics,
  Australian National University, Canberra, ACT 2611, Australia}

\begin{abstract} 

Spectroscopic observations are reported for the 2.75 day,
double-lined, detached eclipsing binary \epic\ located at the turnoff
of the old nearby open cluster \rup. A joint analysis of our
radial velocity measurements and the \ktwo\ light curve leads to
masses of $M_1 = 1.509^{+0.063}_{-0.056}~M_{\sun}$ and $M_2 =
0.649^{+0.015}_{-0.014}~M_{\sun}$ for the primary and secondary, along
with radii of $R_1 = 2.505^{+0.026}_{-0.031}~R_{\sun}$ and $R_2 =
0.652^{+0.013}_{-0.012}~R_{\sun}$, respectively. The effective
temperatures are $6180 \pm 100$~K for the F7 primary and $4010 \pm
170$~K for the late K secondary. The orbit is circular, and the stars'
rotation appears to be synchronized with the orbital motion. This is
the third eclipsing system analyzed in the same cluster, following our
earlier studies of \epicfirst\ and \epicsecond. By comparison with
stellar evolution models from the PARSEC series, we infer an age of
$2.67^{+0.39}_{-0.55}$~Gyr that is consistent with the estimates for
the other two systems. \epic\ is a hierarchical triple system, with
the period of the slightly eccentric outer orbit being 463 days. The
unseen tertiary is either a low-mass M dwarf or a white dwarf.

\end{abstract}

\section{Introduction}
\label{sec:introduction}

Eclipsing binaries that are members of star clusters are particularly
valuable objects for Astrophysics. When they happen to be
double-lined, classical spectroscopic and lightcurve analysis
techniques can yield accurate, model-independent masses and radii with
precisions reaching a few percent in favorable cases \citep[see,
  e.g.,][]{Andersen:1991, Torres:2010}. For detached binaries, such
measurements provide stringent constraints on stellar evolution theory
in a population whose age, metallicity, and distance can be determined
independently based, e.g., on spectroscopic observations and studies
of their color-magnitude diagrams.

In recent work, we reported results for the eclipsing binaries
\epicfirst\ \citep[][hereafter Paper~I]{Torres:2018} and
\epicsecond\ \citep[][Paper II]{Torres:2019} in the nearby old open
cluster \rup\ (NGC~6774). Its members are slightly metal-rich compared
to the Sun \citep[${\rm [Fe/H]} = +0.10$;][]{Curtis:2018}, and are
located some 300~pc away. Both binaries gave consistent ages near
2.7~Gyr from a comparison with models of stellar evolution. While
\epicfirst\ (orbital period $P = 6.53$~days) is composed of very
similar early G-type stars, the components of \epicsecond\ ($P =
11.99$~days) are considerably different in mass (F8 and K5), and
therefore provide greater leverage for testing theory.

In this paper, we present an analysis of a third, short period
(2.75~days) but well detached eclipsing binary system in the same
cluster, \epic. This object was observed by NASA's \ktwo\ mission, the
successor to the \kepler\ mission, during Campaign~7 (late
2015). Aliases include TYC~6296-1893-1, 2MASS~J19162232$-$1627505, and
Gaia DR2 4087856280717586560.  A brief description of the object was
presented by \cite{Curtis:2016}. Its membership in \rup\ is
supported by its position, proper motion, and parallax from the {\it
  Gaia}/DR2 catalog \citep{Gaia:2018}, as reported by
\cite{Olivares:2019}, and also by its systemic radial velocity located
at the peak of the distribution of measures for other cluster members
\citep[see][]{Curtis:2013, Yeh:2019}.  \epic\ is quite bright ($K\!p =
10.12$, $V = 10.34$), and has two distinguishing characteristics: it
is found here to be a triple system for which we are able to determine
the outer orbit, and it is located at the main-sequence turnoff of the
cluster, making its properties (especially the radii) much more
sensitive diagnostics of its evolutionary state. Because of this, it
provides a valuable opportunity to further refine the age
determination for \rup, which can serve as a check on
independent dating techniques such as asteroseismology and
gyrochronology, when those become available for the cluster.

We begin our paper by presenting the photometric, imaging, and
spectroscopic observations in Section~\ref{sec:observations}, along
with the procedures to detrend the photometry and to derive radial
velocities. The joint analysis of the radial velocity measurements and
\ktwo\ light curve is described in Section~\ref{sec:analysis}. The
results are then used to infer the absolute properties of \epic\ in
Section~\ref{sec:dimensions}.  Rotation and activity are discussed in
Section~\ref{sec:rotation}, and Section~\ref{sec:models} deals with a
comparison of the masses, radii, and temperatures against current
models of stellar evolution in order to infer the age. Concluding
remarks are found in Section~\ref{sec:discussion}.

\section{Observations}
\label{sec:observations}

\subsection{Photometry}
\label{sec:photometry}

\epic\ was observed by the \ktwo\ mission during its Campaign~7, as
part of a large super-aperture targeting the core of \rup. The
observations were made in long cadence mode, once every 29.4 minutes.
We downloaded the superstamp observations from the Mikulski Archive
for Space Telescopes (MAST)\footnote{\url{http://archive.stsci.edu/}},
extracted light curves for cluster members following
\cite{Vanderburg:2014} and \cite{Vanderburg:2016}, and initially
identified \epic\ as an eclipsing binary. A total of 30 primary
eclipses and 29 secondary eclipses are included in the 81 days of
photometric coverage. Because the object is in a fairly crowded region
of the sky, we re-extracted a raw light curve following the procedure
from previous papers in our series \citep{Torres:2018, Torres:2019}
using a circular moving aperture with a radius of 15\farcs8, in order
to ensure that the third-light contamination in the lightcurve is
constant (and not dependent on the roll of the \ktwo\ spacecraft).  We
used a first pass systematics correction as in \cite{Vanderburg:2014}
and \cite{Vanderburg:2016}, and then took this as a starting point for
a simultaneous fit of the \ktwo\ 6-hour roll systematics, the primary
and secondary eclipses, and out-of-eclipse variability. Further
details may be found in our previous studies.  We subsequently removed
long term trends with a low-order spline. The photometry processed in
this way is provided in Table~\ref{tab:photometry}, and is used below
for our lightcurve analysis.

\begin{deluxetable}{cc}
\tablewidth{0pt}
\tablecaption{Detrended \ktwo\ Photometry of \epic\ \label{tab:photometry}}
\tablehead{
\colhead{HJD} &
\colhead{}
\\
\colhead{(2,400,000+)} &
\colhead{Residual flux}
}
\startdata
  57301.4866  &  0.99947727 \\
  57301.5070  &  0.99939460 \\
  57301.5275  &  0.99938356 \\
  57301.5479  &  0.99946419 \\
  57301.5683  &  0.99970927
\enddata

\tablecomments{\ktwo\ photometry after removal of instrumental effects
  and long-term drifts. (This table is available in its entirety in
  machine-readable form.)}

\end{deluxetable}

\subsection{Spectroscopy}
\label{sec:spectroscopy}

\epic\ was monitored spectroscopically at the Center for Astrophysics
for three years beginning in 2016 September, with the fiber-fed,
bench-mounted Tillinghast Reflector Echelle Spectrograph
\citep[TRES;][]{Szentgyorgyi:2007, Furesz:2008} attached to the 1.5m
Tillinghast reflector at the Fred L.\ Whipple Observatory on Mount
Hopkins (Arizona, USA).  We collected a total of 43 spectra, at a
resolving power of $R \approx 44,000$ and covering the wavelength
region 3800--9100~\AA\ in 51 orders. For the order centered at
$\sim$5187~\AA\ containing the \ion{Mg}{1}~b triplet, the
signal-to-noise ratios range from 46 to 100 per resolution element of
6.8~\kms.

While visual examination shows the spectra to be only single-lined,
the radial velocity of the very weak secondary lines can be measured
in most cases using TODCOR, a two-dimensional cross-correlation
technique introduced by \cite{Zucker:1994}. Templates matching the
properties of each component were taken from a pre-computed library of
synthetic spectra that are based on model atmospheres by
R.\ L.\ Kurucz, and a line list tuned to better match the spectra of
real stars \citep[see][]{Nordstrom:1994, Latham:2002}. These templates
cover a limited wavelength region of $\sim$300~\AA\ centered around
5187~\AA.

The effective temperature ($T_{\rm eff}$) and projected rotational
velocity ($v \sin i$) of the primary star were determined following
the procedure described by \cite{Torres:2002}, by running grids of
one-dimensional cross-correlations of the observed spectra against
synthetic spectra over broad ranges in those two parameters. We
ignored the presence of the faint secondary, as it does not affect the
results.  We then selected the combination of parameters giving the
highest value of the cross-correlation coefficient averaged over all
43 spectra, weighted by the strength of each exposure. We repeated
this for fixed values of the surface gravity ($\log g$) of 3.5 and
4.0, bracketing the final values reported below in
Section~\ref{sec:dimensions}, and for metallicities [Fe/H] of 0.0 and
+0.5 on either side of the known cluster abundance. By interpolation
we obtained $T_{\rm eff} = 6180$~K and $v \sin i = 49~\kms$, with
estimated uncertainties of 100~K and 3~\kms, respectively. These
errors are based on the scatter from the individual spectra,
conservatively increased to account for possible systematic
errors. The corresponding spectral type for the primary is
approximately F7. For the radial velocity determinations of this star,
we used template parameters of 6250~K and 50~\kms, which are the
nearest in our grid, along with $\log g = 4.0$ and ${\rm [Fe/H]} =
0.0$.

The lines of the secondary are too weak for us to determine its
temperature or its rotational velocity directly from our spectra.  For
the purpose of measuring radial velocities, we therefore used a
template with $T_{\rm eff} = 4000$~K appropriate for star of its mass
as determined later (spectral type late K). In
Section~\ref{sec:dimensions} below we provide an empirical estimate of
the secondary temperature that supports this choice. For the rotation,
we adopted $v \sin i = 12$~\kms\ assuming that its spin is
synchronized with the orbital motion, and using a typical radius for a
star of this type. The metallicity for the secondary template was kept
at the solar value as for the primary, and $\log g$ was set to 4.5.

Each of the 43 spectra yielded a precise radial velocity
measurement for the primary, but the secondary lines were clearly
visible in only 31 of them. The heliocentric velocities for both stars
in \epic\ are presented in Table~\ref{tab:rvs}, along with their
formal uncertainties. The secondary velocities are much poorer because
of its faintness. Using TODCOR we estimated the average
secondary-to-primary flux ratio at the mean wavelength of our
observations (5187~\AA) to be only $\ell_2/\ell_1 = 0.0035 \pm
0.0010$, the smallest we have ever measured with this instrument for
any star.

\setlength{\tabcolsep}{3.5pt}  
\begin{deluxetable}{c@{}c@{}c@{}cc}
\tablewidth{0pc}
\tablecaption{Heliocentric Radial-velocity Measurements of \epic \label{tab:rvs}}
\tablehead{
\colhead{HJD} &
\colhead{RV$_1$} &
\colhead{RV$_2$} &
\colhead{Inner} &
\colhead{Outer}
\\
\colhead{(2,400,000$+$)} &
\colhead{(\kms)} &
\colhead{(\kms)} &
\colhead{Phase} &
\colhead{Phase}
}
\startdata
 57647.6200  &    $-20.98 \pm 0.38$\phs  &         \nodata               &  0.7552  &   0.0738  \\      
 57853.9799  &    $-13.71 \pm 0.33$\phs  &  $182.93 \pm 11.51$           &  0.6978  &   0.5191  \\      
 57857.9907  &    $ 90.89 \pm 0.27$      &  $-62.88 \pm  9.39$\phs       &  0.1544  &   0.5277  \\      
 57863.9704  &    $ 94.87 \pm 0.31$      &  \phn$ -86.73 \pm 10.89$\phs  &  0.3260  &   0.5406  \\      
 57878.9624  &    $-15.33 \pm 0.22$\phs  &  $169.53 \pm  7.67$\phn       &  0.7706  &   0.5730  \\      
 57879.9634  &     $87.33 \pm 0.21$      &  $-56.32 \pm  7.46$\phs       &  0.1341  &   0.5751  \\      
 57885.9000  &    $100.55 \pm 0.23$\phn  &         \nodata               &  0.2901  &   0.5879  \\      
 57900.9030  &    $-14.16 \pm 0.23$\phs  &         \nodata               &  0.7386  &   0.6203  \\      
 57906.8573  &    $ 10.28 \pm 0.20$      &         \nodata               &  0.9010  &   0.6331  \\      
 57907.9249  &    $101.66 \pm 0.20$\phn  &  $-92.58 \pm  7.01$\phs       &  0.2887  &   0.6354  \\      
 57908.8536  &  \phn$2.35 \pm 0.23$      &  $128.76 \pm  8.08$\phn       &  0.6260  &   0.6374  \\      
 57910.8995  &     $88.14 \pm 0.33$      &   \phn$-69.46 \pm 11.54$\phs  &  0.3690  &   0.6419  \\      
 57914.9432  & \phn$-4.84 \pm 0.26$\phs  &  $165.37 \pm  9.22$\phn       &  0.8375  &   0.6506  \\      
 57919.8292  &  \phn$6.84 \pm 0.22$      &  $144.13 \pm  7.79$\phn       &  0.6120  &   0.6611  \\      
 57932.9240  &     $88.91 \pm 0.25$      &  $-45.51 \pm  8.60$\phs       &  0.3675  &   0.6894  \\      
 57965.7724  &    $102.98 \pm 0.26$\phn  &  $-98.09 \pm  9.04$\phs       &  0.2969  &   0.7603  \\      
 58001.6828  &     $96.93 \pm 0.26$      &         \nodata               &  0.3383  &   0.8377  \\      
 58002.6854  &    $-10.32 \pm 0.24$\phs  &  $170.88 \pm  8.58$\phn       &  0.7024  &   0.8399  \\      
 58020.6379  &    $104.28 \pm 0.23$\phn  &         \nodata               &  0.2221  &   0.8786  \\      
 58034.6044  &    $101.65 \pm 0.20$\phn  &         \nodata               &  0.2943  &   0.9088  \\      
 58035.5836  & \phn$-2.20 \pm 0.23$\phs  &  $151.53 \pm  8.08$\phn       &  0.6499  &   0.9109  \\      
 58037.5873  &     $86.33 \pm 0.24$      &  $-45.23 \pm  8.51$\phs       &  0.3775  &   0.9152  \\      
 58050.5881  &     $78.15 \pm 0.25$      &  $-38.64 \pm  8.68$\phs       &  0.0990  &   0.9433  \\      
 58056.5695  &    $102.00 \pm 0.24$\phn  &  $-93.30 \pm  8.34$\phs       &  0.2712  &   0.9562  \\      
 58060.5644  &    $-15.34 \pm 0.22$\phs  &  $181.35 \pm  7.83$\phn       &  0.7220  &   0.9648  \\      
 58068.5504  &  \phn$0.61 \pm 0.28$      &  $134.30 \pm  9.66$\phn       &  0.6222  &   0.9820  \\      
 58259.9510  &     $82.52 \pm 0.22$      &  $-51.94 \pm  7.60$\phs       &  0.1321  &   0.3950  \\      
 58276.8394  &     $99.05 \pm 0.38$      &         \nodata               &  0.2654  &   0.4314  \\      
 58277.9011  & \phn$-8.36 \pm 0.24$\phs  &  $147.55 \pm  8.36$\phn       &  0.6510  &   0.4337  \\      
 58279.8654  &     $84.53 \pm 0.21$      &         \nodata               &  0.3644  &   0.4380  \\      
 58290.8725  &     $85.49 \pm 0.19$      &  $-52.88 \pm  6.69$\phs       &  0.3618  &   0.4617  \\      
 58301.8394  &     $89.74 \pm 0.21$      &  $-85.32 \pm  7.44$\phs       &  0.3445  &   0.4854  \\      
 58331.9025  &    $101.77 \pm 0.42$\phn  &  \phn$-95.65 \pm 14.71$\phs   &  0.2624  &   0.5502  \\      
 58386.7016  &     $95.58 \pm 0.27$      &  $-66.91 \pm  9.32$\phs       &  0.1635  &   0.6685  \\      
 58594.9671  &    $-19.88 \pm 0.31$\phs  &         \nodata               &  0.7981  &   0.1178  \\      
 58601.9686  &     $86.14 \pm 0.25$      &  $-71.51 \pm  8.85$\phs       &  0.3408  &   0.1329  \\      
 58621.9451  &  \phn$2.58 \pm 0.27$      &  $122.40 \pm  9.59$\phn       &  0.5956  &   0.1760  \\      
 58628.8947  &     $75.78 \pm 0.21$      &  $-49.69 \pm  7.28$\phs       &  0.1194  &   0.1910  \\      
 58634.9162  &     $91.20 \pm 0.23$      &  $-88.82 \pm  8.10$\phs       &  0.3062  &   0.2040  \\      
 58660.8715  &    $-21.93 \pm 0.27$\phs  &  $172.06 \pm  9.36$\phn       &  0.7323  &   0.2600  \\      
 58674.8176  &    $-19.51 \pm 0.30$\phs  &  $169.69 \pm 10.50$           &  0.7970  &   0.2901  \\      
 58693.7778  &    $-15.90 \pm 0.30$\phs  &         \nodata               &  0.6827  &   0.3310  \\      
 58744.6811  &    $ 91.77 \pm 0.34$      &         \nodata               &  0.1690  &   0.4408          
\enddata

\tablecomments{Orbital phases for the inner orbit are counted from the
  reference time of primary eclipse, and those for the outer orbit
  from the corresponding time of periastron passage. The final
  velocity uncertainties for our analysis result from scaling the
  values listed here for the primary and secondary by the near-unity
  factors $f_1$ and $f_2$, respectively, from our global analysis
  described in Section~\ref{sec:analysis}.}

\end{deluxetable}
\setlength{\tabcolsep}{6pt}  

A preliminary orbital solution based on these velocities showed an
obvious long-term periodic pattern in the residuals of the primary
star, with a peak-to-peak amplitude of about 10~\kms. This indicates
the presence of a third component in the system. However, careful
examination of our spectra with TRICOR, an extension of TODCOR to
three dimensions \citep{Zucker:1995}, showed no sign of a third set of
lines. This suggests that the tertiary must be even fainter than the
secondary, possibly a mid or late M dwarf.

A preliminary fit to the velocities was performed to serve as a
starting point for the analysis of Section~\ref{sec:analysis}, solving
for the elements of the inner and outer orbits simultaneously assuming
they are represented by independent Keplerian trajectories. The outer
orbit is slightly eccentric ($e \approx 0.19$) and has a period of
about 463 days that is covered more than twice by our observations.
The inner orbit for the eclipsing pair, on the other hand, shows no
significant eccentricity.

\begin{figure}
\epsscale{1.15}
\plotone{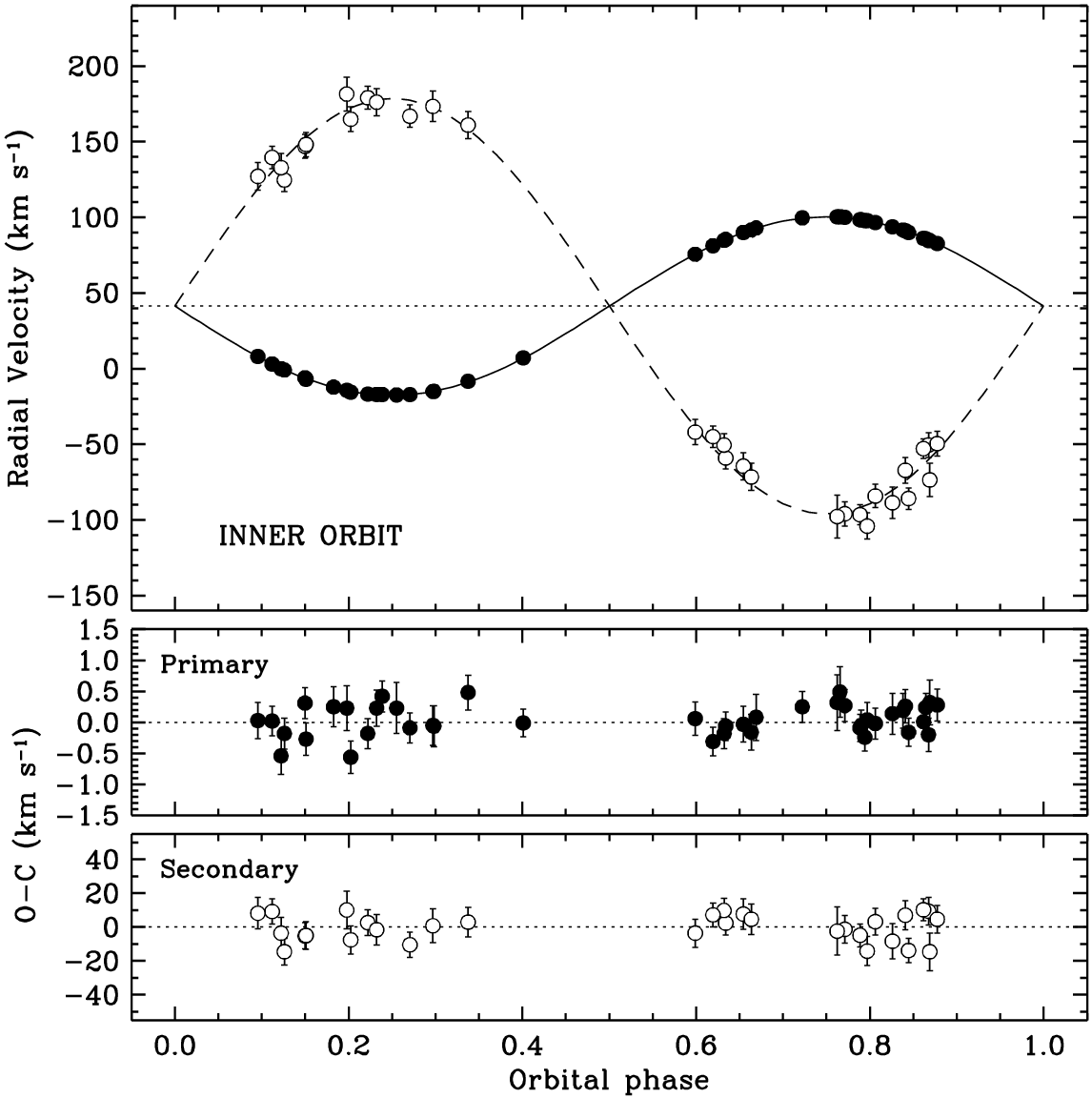}

\figcaption{Radial-velocity measurements for \epic, with our adopted
  model for the inner orbit from Section~\ref{sec:analysis}. Primary
  and secondary measurements are represented with filled and open
  circles, respectively, and have the motion in the outer orbit
  removed. The dotted line marks the center-of-mass velocity of the
  triple system. Error bars for the primary are too small to be
  visible. They are seen in the lower panels, which display the
  residuals. Phases are counted from the reference time of primary
  eclipse (Table~\ref{tab:mcmc}).\label{fig:rvs_inner}}

\end{figure}

\begin{figure}
\epsscale{1.15}
\plotone{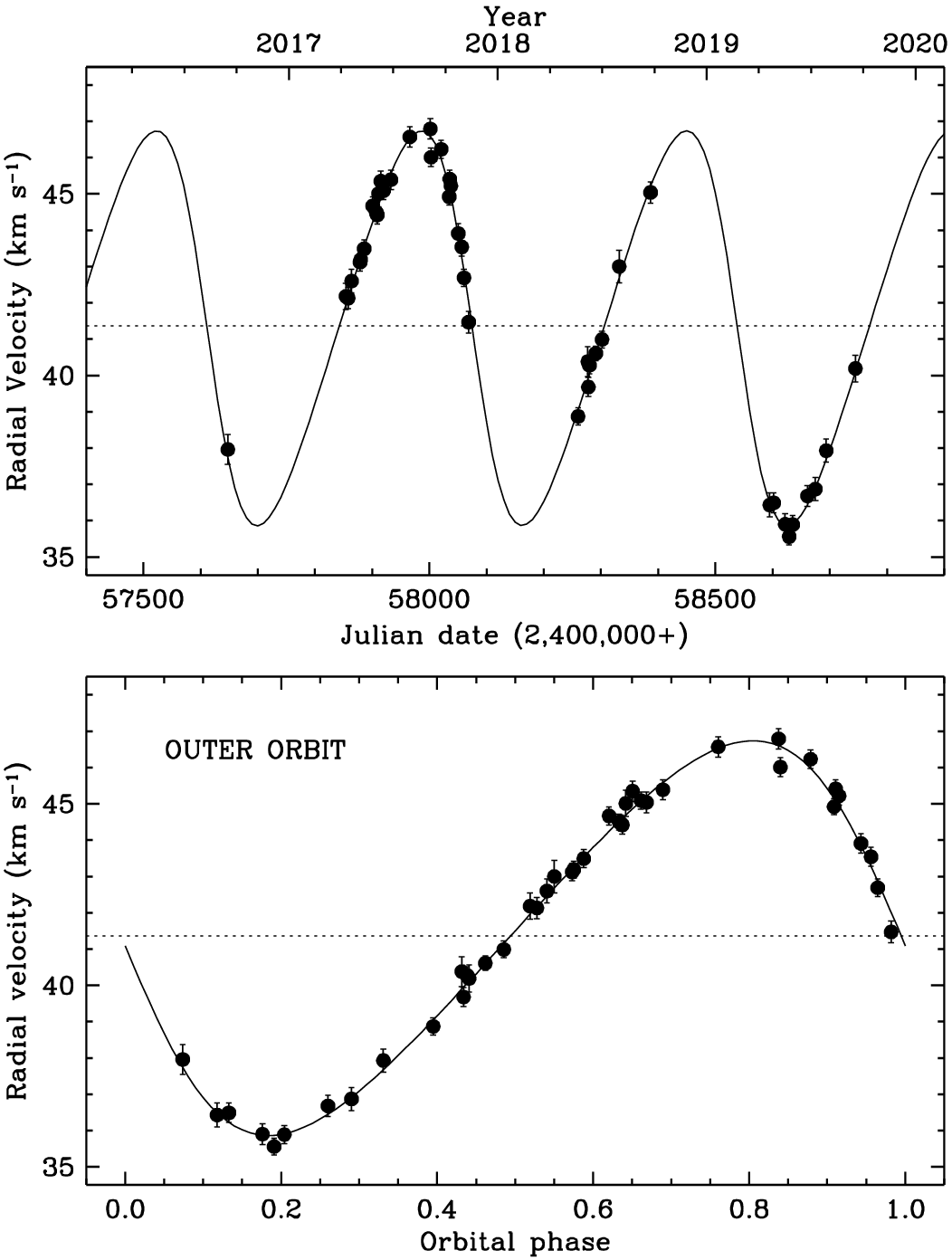}

\figcaption{Radial-velocity measurements for the primary of \epic\ in
  the outer orbit, as a function of time (top) and orbital phase
  counted from periastron passage (bottom). Motion in the inner orbit
  has been removed. The secondary has much larger scatter and is not
  shown, for clarity. The solid line is our final model from
  Section~\ref{sec:analysis}, and the dotted line represents the
  center-of-mass velocity of the triple. \label{fig:rvs_outer}}

\end{figure}

Figure~\ref{fig:rvs_inner} displays the velocities of the primary and
secondary in the inner (2.75~day) orbit after subtracting the motion
in the outer orbit, as described below. Our final model is also shown.
The motion of the primary star in the outer orbit is illustrated in
Figure~\ref{fig:rvs_outer}, in which we have removed the short-period
motion in the inner orbit.

\subsection{Imaging}
\label{sec:imaging}

\begin{figure}
\epsscale{1.15}
\includegraphics[trim= 20pt 10pt 0pt 35pt,width=0.5\textwidth]{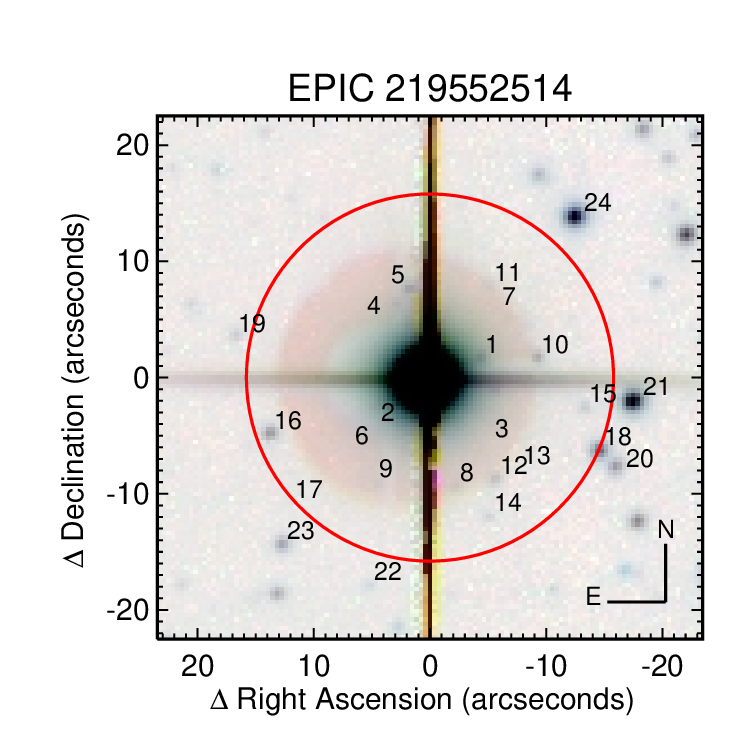}

\figcaption{CFHT $r$-band image of the field of \epic, with the
  15\farcs8 photometric aperture used to extract the
  \ktwo\ photometry indicated with a circle. Nearby companions are
  numbered as in Table~\ref{tab:CFHT}.\label{fig:CFHT}}

\end{figure}

The aperture we used to extract the photometry of \epic\ appears
fairly clear of any intruding stars bright enough to add significant
flux to the light curve and bias the results of our analysis
below. This is shown in Figure~\ref{fig:CFHT}, which is a
seeing-limited image in a bandpass similar to Sloan $r$ (close to
\kepler's $K\!p$ bandpass) taken in 2008 by \cite{Curtis:2013} with
the MegaCam instrument \citep{Hora:1994} on the Canada-France-Hawaii
Telescope (CFHT). The positions of all numbered stars in or near the
aperture, their separations $\rho$ from the target, and their
brightness in the CFHT $gri$
filters\footnote{\url{http://www.cadc-ccda.hia-iha.nrc-cnrc.gc.ca/en/megapipe/docs/filtold.html}},
are given in Table~\ref{tab:CFHT} when bright enough to measure. We
include also $J$- and $K$-band brightness measurements based on
UKIRT/WFCAM imaging \citep{Curtis:2016}, which reaches deeper. We
additionally report the $G$-band magnitude and trigonometric parallax,
when available, for the few companions that have entries in the {\it
  Gaia}/DR2 catalog \citep{Gaia:2018}. None appear to be members of
the cluster. All companions within the aperture are very faint and
have no effect on our analysis. Even the two brighter ones that are
slightly outside the aperture (\#21 and \#24) will not contribute
significantly: they are more than 6 magnitudes fainter than the target
in the near infrared, they are fairly red (SpT $\sim$ K3 and K5,
respectively) and will therefore appear even fainter in the $K\!p$
band, and only a small fraction of their flux would be inside the
aperture given the \kepler\ pixel scale of $3\farcs98$~pix$^{-1}$.

\setlength{\tabcolsep}{4.5pt}
\begin{deluxetable*}{rccccccccccccc}
\tablewidth{0pt}
\tablecaption{Close Neighbors of \epic\ \label{tab:CFHT}}
\tablehead{
\colhead{} &
\colhead{R.A.} &
\colhead{Dec.} &
\colhead{P.A.} &
\colhead{$\rho$} & 
\colhead{$J$} &
\colhead{$K$} &
\colhead{$\sigma_{JK}$} &
\colhead{$g$} &
\colhead{$r$} &
\colhead{$i$} &
\colhead{$\sigma_{gri}$} & 
\colhead{$G$} &
\colhead{$\pi_{Gaia}$}
\\
\colhead{\#} &
\colhead{(J2000)} &
\colhead{(J2000)} &
\colhead{(degree)} &
\colhead{($\arcsec$)} &
\colhead{(mag)} &
\colhead{(mag)} &
\colhead{(mag)} &
\colhead{(mag)} &
\colhead{(mag)} &
\colhead{(mag)} &
\colhead{(mag)} &
\colhead{(mag)} &
\colhead{(mas)}
}
\startdata
 1 & 19:16:22.01 & $-$16:27:48.9 & 295.2       & \phn5.2 & 17.05 & 16.28 & 0.02 & \nodata & \nodata & \nodata & \nodata & \nodata & \nodata          \\  
 2 & 19:16:22.61 & $-$16:27:54.8 & 132.4       & \phn5.7 & 17.66 & 16.65 & 0.02 & \nodata & \nodata & \nodata & \nodata & \nodata & \nodata          \\  
 3 & 19:16:21.96 & $-$16:27:56.2 & 226.1       & \phn7.6 & 18.67 & 17.54 & 0.05 & \nodata & \nodata & \nodata & \nodata & \nodata & \nodata          \\  
 4 & 19:16:22.69 & $-$16:27:45.6 & \phn43.5    & \phn7.7 & 18.52 & 17.51 & 0.04 & \nodata & \nodata & \nodata & \nodata & \nodata & \nodata          \\  
 5 & 19:16:22.42 & $-$16:27:43.0 & \phn\phn9.3 & \phn8.2 & 18.07 & 17.44 & 0.04 & \nodata & \nodata & \nodata & \nodata & 20.72   & \nodata          \\  
 6 & 19:16:22.76 & $-$16:27:56.8 & 132.8       & \phn8.6 & 18.82 & 19.00 & 0.17 & \nodata & \nodata & \nodata & \nodata & \nodata & \nodata          \\  
 7 & 19:16:21.92 & $-$16:27:44.9 & 316.4       & \phn8.8 & 18.94 & 18.35 & 0.09 & \nodata & \nodata & \nodata & \nodata & \nodata & \nodata          \\  
 8 & 19:16:22.16 & $-$16:27:60.0 & 195.8       & \phn9.3 & 19.47 & 18.59 & 0.12 & \nodata & \nodata & \nodata & \nodata & \nodata & \nodata          \\  
 9 & 19:16:22.63 & $-$16:27:59.6 & 153.6       & \phn9.6 & 19.30 & 18.02 & 0.08 & \nodata & \nodata & \nodata & \nodata & \nodata & \nodata          \\  
10 & 19:16:21.69 & $-$16:27:49.0 & 283.0       & \phn9.8 & 19.06 & 18.61 & 0.12 & \nodata & \nodata & \nodata & \nodata & 20.80   & \nodata          \\  
11 & 19:16:21.96 & $-$16:27:42.7 & 327.6       & 10.0    & 19.43 & 19.02 & 0.17 & \nodata & \nodata & \nodata & \nodata & \nodata & \nodata          \\  
12 & 19:16:21.93 & $-$16:27:59.3 & 214.9       & 10.2    & 20.12 & 19.72 & 0.32 & \nodata & \nodata & \nodata & \nodata & \nodata & \nodata          \\  
13 & 19:16:21.80 & $-$16:27:58.5 & 226.0       & 10.9    & 20.44 & 18.88 & 0.23 & \nodata & \nodata & \nodata & \nodata & \nodata & \nodata          \\  
14 & 19:16:21.96 & $-$16:28:02.6 & 204.7       & 12.8    & 19.83 & 19.43 & 0.25 & \nodata & \nodata & \nodata & \nodata & \nodata & \nodata          \\  
15 & 19:16:21.42 & $-$16:27:53.2 & 260.8       & 13.8    & 19.57 & 19.19 & 0.20 & 23.29   & 22.41   & 21.99   & 0.07    & \nodata & \nodata          \\  
16 & 19:16:23.22 & $-$16:27:55.5 & 108.9       & 14.1    & 18.67 & 18.34 & 0.09 & 20.83   & 20.40   & 20.23   & 0.14    & 20.19   & $1.68 \pm 0.80$  \\  
17 & 19:16:23.11 & $-$16:28:01.4 & 132.8       & 15.5    & 19.60 & 17.97 & 0.11 & \nodata & \nodata & \nodata & \nodata & \nodata & \nodata          \\  
18 & 19:16:21.34 & $-$16:27:56.9 & 247.8       & 16.0    & 17.43 & 16.82 & 0.02 & 20.04   & 19.26   & 18.91   & 0.02    & 19.24   & $0.63 \pm 0.37$  \\  
19 & 19:16:23.43 & $-$16:27:47.2 & \phn76.1    & 16.9    & 19.55 & 18.69 & 0.13 & 23.53   & 22.55   & 22.30   & 0.07    & \nodata & \nodata          \\  
20 & 19:16:21.24 & $-$16:27:58.2 & 245.5       & 17.9    & 18.55 & 18.00 & 0.07 & 20.90   & 20.13   & 19.91   & 0.02    & 20.25   & $0.48 \pm 0.88$  \\  
21 & 19:16:21.14 & $-$16:27:52.6 & 264.9       & 17.9    & 15.86 & 15.30 & 0.01 & \nodata & \nodata & \nodata & \nodata & 17.44   & $-0.11 \pm 0.12$\phs \\  
22 & 19:16:22.65 & $-$16:28:08.5 & 165.3       & 18.1    & 19.63 & 19.28 & 0.21 & \nodata & \nodata & \nodata & \nodata & \nodata & \nodata          \\  
23 & 19:16:23.15 & $-$16:28:05.0 & 139.6       & 18.6    & 18.47 & 17.79 & 0.06 & 21.52   & 20.58   & 20.27   & 0.03    & 20.43   & $0.90 \pm 0.90$  \\  
24 & 19:16:21.48 & $-$16:27:36.7 & 319.5       & 19.2    & 15.82 & 15.14 & 0.01 & 18.54   & 17.63   & 17.28   & 0.01    & 17.65   & $0.27 \pm 0.15$    
\enddata

\tablecomments{Coordinates based on UKIRT images
  \citep[see][]{Curtis:2016}. Average uncertainties $\sigma_{JK}$ and
  $\sigma_{gri}$ are listed for the corresponding magnitude
  measurements.}

\end{deluxetable*}

In order to search for blended stellar companions to \epic\ inside the
inner working angle of the seeing-limited imaging, we used the 10m
Keck~II telescope with the NIRC2 facility adaptive optics (AO) imager
to obtain natural guide star adaptive optics imaging and non-redundant
aperture mask interferometry (NRM). These observations were made in
the $K'$ filter ($\lambda = 2.124~\mu$m) on 2016 June 16 UT, and
followed the standard observing strategy described by
\cite{Kraus:2016} and previously reported for \rup\ targets by
\cite{Torres:2018} and \cite{Torres:2019}.  For \epic, we obtained a
short sequence of 6 images and 8 interferograms in vertical angle
mode. In both cases, calibrators were drawn from the other
\rup\ members observed on the same night.

The images were analyzed following the methods described by
\cite{Kraus:2016}. To summarize, the primary star point spread
function (PSF) was subtracted using both an azimuthal median profile
and the calibrator that most closely matches the speckle
pattern. Within each image, the residual fluxes as a function of
position were measured in apertures of radius 40 milli-arc seconds
(mas), centered on each pixel, and the noise was estimated from the
RMS of fluxes within concentric rings around the primary
star. Finally, the detections and detection limits were estimated from
the flux-weighted sum of the detection significances in the stack of
all images, and any location with a total significance greater than
6$\sigma$ was visually inspected to determine if it was a residual
speckle or cosmic ray. No candidates remained after this visual
inspection. The observations yielded contrast limits of $\Delta K' =
5.6$~mag at $\rho = 150$~mas, $\Delta K' = 7.7$~mag at $\rho =
500$~mas, and $\Delta K' = 9.0$~mag at $\rho > 1000$~mas.

The interferograms were analyzed following the methods described by
\cite{Kraus:2008} and \cite{Ireland:2013}. We Fourier-transformed the
interferograms to extract the complex visibilities, and from those we
computed the corresponding closure phases for each triplet of
baselines. We calibrated the closure phases with other observations of
targets nearby on the sky and in time, and then fit the calibrated
closure phases with binary source models to search for significant
evidence of a companion, but did not find any. We determined the
detection limits using a Monte Carlo process that randomizes the phase
errors and determines the distribution of possible binary fits,
indicating the 99.9\% upper limit on companions in bins of projected
separation.  The observations yielded contrast limits of $\Delta K' =
0.13$~mag at $\rho = 20$--40~mas, $\Delta K' = 1.22$~mag at $\rho =
40$--80~mas, and $\Delta K' = 0.73$~mag at $\rho = 80$--160~mas.

\section{Analysis}
\label{sec:analysis}

For the analysis of the \ktwo\ light curve, we adopted the
Nelson-Davis-Etzel binary model \citep{Etzel:1981, Popper:1981} as
implemented in the {\tt eb} code of \cite{Irwin:2011}.  This model
approximates the star shapes as biaxial spheroids for calculating
proximity effects, and is adequate for well-detached systems in which
the stars are nearly spherical, as is the case here (see below).  The
main adjustable parameters we considered for the inner binary are as
follows: the orbital period ($P_{\rm in}$), a reference epoch of
primary eclipse ($T_0$, which is strictly the time of inferior
conjunction in this code), the central surface brightness ratio in the
\kepler\ bandpass ($J \equiv J_2/J_1$), the sum of the relative radii
normalized by the semimajor axis ($r_1+r_2$) and their ratio ($k
\equiv r_2/r_1$), the cosine of the inclination angle ($\cos i$), the
eccentricity parameters $e_{\rm in} \cos\omega_{\rm in}$ and $e_{\rm
  in} \sin\omega_{\rm in}$, with $e_{\rm in}$ being the eccentricity
and $\omega_{\rm in}$ the longitude of periastron for the primary, and
an out-of-eclipse brightness level in magnitude units ($m_0$).  We
adopted a quadratic limb-darkening law for this work, with
coefficients $u_1$ and $u_2$ for the primary and a corresponding set
for the secondary. The reflection albedos ($A_1$, $A_2$) were included
as additional variables.  Gravity darkening coefficients for the
\kepler\ band were adopted from the theoretical calculations by
\cite{Claret:2011}, interpolated to the metallicity of \rup,
the temperatures indicated earlier, and the final $\log g$ values
reported below. They were held fixed at the values $y_1 = 0.306$ for
the primary and $y_2 = 0.460$ for the secondary.\footnote{Note that
  these are bandpass-specific \emph{coefficients}, not to be confused
  with the bolometric gravity darkening \emph{exponents} used in other
  eclipsing binary modeling programs \citep[see][]{Torres:2017}.}

Even though there is no evidence of significant flux from neighboring
stars in the photometric aperture, as a precaution we included the
third light parameter $\ell_3$ as an additional adjustable parameter
because the unseen tertiary (presumably a very red star) may be
brighter in the \kepler\ band (centered around 6000~\AA) than in our
spectroscopic window ($\sim$5187~\AA).  Third light is defined here
such that $\ell_1 + \ell_2 + \ell_3 = 1$, and the values for the
primary and secondary for this normalization correspond to the light
at first quadrature.

To avoid biases, the finite integration time of the
\ktwo\ long-cadence observations was accounted for by oversampling the
model light curve and then integrating over the 29.4-minute duration
of each cadence prior to the comparison with the observations
\citep[see][]{Gilliland:2010, Kipping:2010}.

The radial velocities were included in the analysis along with the
photometry, and the spectroscopic elements for the inner and outer
orbits were solved simultaneously as done in
Section~\ref{sec:spectroscopy}.  This introduces the following
additional elements: the primary and secondary velocity semiamplitudes
($K_1$ and $K_2$), the center-of-mass velocity of the triple system
($\gamma$), the outer orbital period ($P_{\rm out}$), a reference time
of periastron passage for the outer orbit ($T_{\rm peri}$), the
velocity semiamplitude of the inner binary in the outer orbit ($K_{\rm
  out}$), and the eccentricity parameters $e_{\rm out} \cos\omega_{\rm
  out}$ and $e_{\rm out} \sin\omega_{\rm out}$, where $\omega_{\rm
  out}$ corresponds to the longitude of periastron of the inner
binary. We point out here that a mismatch between our
cross-correlation templates and the real stars can potentially
introduce a spurious systematic offset between the measured primary
and secondary velocities.  Because the template parameters adopted for
the secondary, particularly $v \sin i$, are merely educated guesses,
we allowed for such an offset ($\Delta$RV) that we added to the list
of free parameters.

Light travel time in the outer orbit will cause the eclipses to arrive
slightly earlier or later than they would in the absence of the
tertiary. Over the $\sim$80 days of the \ktwo\ observations the effect
varies between $-1.6$ and $-2.3$ minutes (eclipses occur earlier),
which is significant compared to the final precision we report for
$T_0$ below. Consequently, we accounted for this effect during the
analysis. This was done by appropriately adjusting all times of
observation based on a estimate of those corrections from a
preliminary model that used the radial velocities alone.

Our method of solution used the {\tt
  emcee\/}\footnote{\url{http://dan.iel.fm/emcee}} code of
\cite{Foreman-Mackey:2013}, which is a Python implementation of the
affine-invariant Markov chain Monte Carlo (MCMC) ensemble sampler
proposed by \cite{Goodman:2010}. We used 100 walkers with chain
lengths of 15,000 each, after discarding the burn-in.  Uniform
(non-informative) or log-uniform priors over suitable ranges were
adopted for most adjustable parameters (see below), and convergence
was verified by examining the chains visually and by requiring a
Gelman-Rubin statistic of 1.05 or smaller for each parameter
\citep{Gelman:1992}.  For more efficient sampling of parameter space,
and to reduce the correlation between them, the traditional quadratic
limb-darkening coefficients $u_1$ and $u_2$ for each star were recast
as $q_1$ and $q_2$ following \cite{Kipping:2013}, where $q_1 = (u_1 +
u_2)^2$ and $q_2 = 0.5 u_1/(u_1+u_2)$.

\begin{figure}
\epsscale{1.15}
\includegraphics[width=0.478\textwidth]{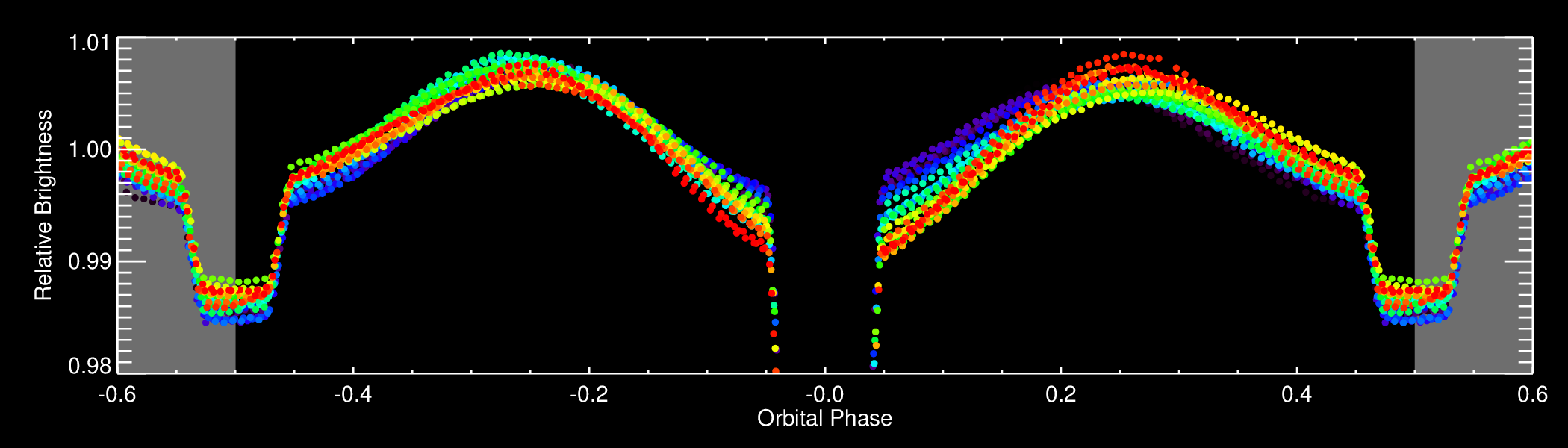}
\includegraphics[width=0.478\textwidth]{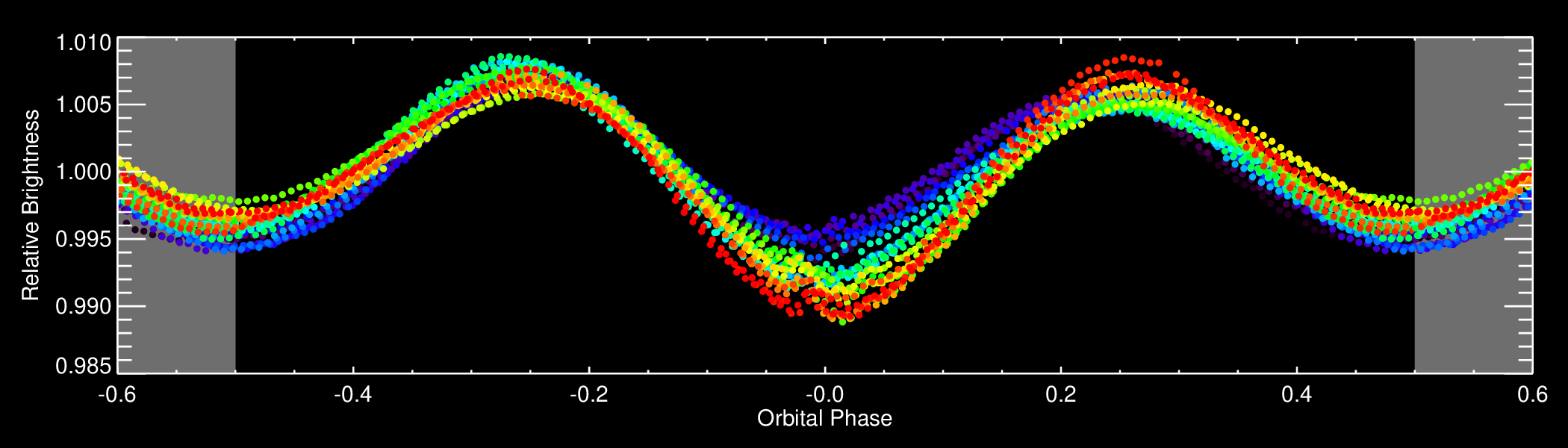}

\figcaption{Phase-folded photometry of \epic, color-coded by date to
  show the evolution of lightcurve distortions presumably caused by
  spots. Red points correspond to earlier data, and purple points to
  later data. The bottom panel has the eclipses removed, and reveals
  what appears to be a spot-crossing event (bump) during the primary
  eclipse in the early observations (red), which gradually disappears
  and is no longer seen in the later data (purple). \label{fig:spots}}

\end{figure}

The relative weighting between the photometry and radial velocity
measurements was handled by introducing additional free parameters in
the form of multiplicative scale factors for the observational
errors. These scale factors ($f_{\ktwo}$ for the photometry, and $f_1$
and $f_2$ for the primary and secondary velocities) were solved for
self-consistently and simultaneously with the other orbital quantities
\citep[see][]{Gregory:2005}. The initial error assumed for the
photometric measurements is 1.6 milli-magnitudes (mmag), which is
approximately the out-of-eclipse scatter, and the initial errors for
the velocities are those listed in Table~\ref{tab:rvs}. The large
scatter in the phase-folded photometry compared to the typical
precision of the \ktwo\ instrument (roughly 50 parts per million per
half-hour integration for a non-variable star of this brightness) is
caused by obvious distortions presumably due to spots, which appear to
be changing on very short timescales of days. This is illustrated in
Figure~\ref{fig:spots}, and discussed further below.

Initial tests showed that the second-order limb-darkening coefficients
$q_2$ were essentially unconstrained for both stars, likely because of
the light curve distortions just mentioned. We therefore held those
coefficients fixed at their theoretical values for the \kepler\ band
according to \cite{Claret:2011}, selecting the ones based on ATLAS
model atmospheres and the least-squares fitting procedure favored by
those authors.  The tabulated $u_2$ values in the standard quadratic
limb-darkening formulation are 0.305 and 0.182 for the primary and
secondary.  For the linear coefficients $q_1$ we adopted Gaussian
priors from theory with a standard deviation of 0.1 (see
Table~\ref{tab:mcmc}). We also found that while the albedo for the
secondary was well constrained, the one for the primary was not. We
chose to impose weak Gaussian priors on both, with a mean of 0.5
(appropriate for convective stars) and a standard deviation of
0.3. Finally, all our tests indicated a negligible eccentricity for
the inner orbit, consistent with the findings from the spectroscopy.
To reduce the already large number of free parameters, we set the
eccentricity to zero for the remainder of this work.

The results of our analysis for \epic\ are reported in
Table~\ref{tab:mcmc}, in which the values given correspond to the mode
of the posterior distributions. The distributions of the derived
quantities listed in the bottom section of the table were constructed
directly from the MCMC chains of the adjustable parameters
involved. Included among these is $J_{\rm ave}$, the surface
brightness ratio averaged over the stellar disk, and the flux ratio
$\ell_2/\ell_1$ in the \kepler\ band.  Both stars in the inner binary
are found to be nearly spherical, justifying the use of this binary
model.  We calculate the oblateness of the primary star as defined by
\cite{Binnendijk:1960} to be 0.008, which is well below the safe limit
for this binary model \citep[0.04; see, e.g.,][]{Popper:1981}. The
oblateness of the secondary is an order of magnitude smaller.

\setlength{\tabcolsep}{2.5pt}
\begin{deluxetable}{lcc}
\tablewidth{0pc}
\tablecaption{Results from our Combined MCMC Analysis for \epic \label{tab:mcmc}}
\tablehead{ \colhead{~~~~~~~~~~~Parameter~~~~~~~~~~~} & \colhead{Value} & \colhead{Prior} }
\startdata
 $P_{\rm in}$ (days)\dotfill      &  $2.7535724^{+0.0000019}_{-0.0000024}$ & [2, 3] \\ [1ex]
 $T_0$ (HJD$-$2,400,000)\dotfill  &  $57357.79215^{+0.00019}_{-0.00018}$   & [57356, 57358] \\ [1ex]
 $J$\dotfill                      &  $0.107^{+0.014}_{-0.014}$             & [0.02, 1.0] \\ [1ex]
 $r_1+r_2$\dotfill                &  $0.2949^{+0.0011}_{-0.0010}$          & [0.01, 0.50] \\ [1ex]
 $k$\dotfill                      &  $0.2603^{+0.0038}_{-0.0035}$          & [0.1, 1.0] \\ [1ex]
 $\cos i$\dotfill                 &  $0.002^{+0.018}_{-0.002}$             & [0, 1] \\ [1ex]
 $m_0$ (mag)\dotfill              &  $10.11773^{+0.00030}_{-0.00032}$      & [9, 11] \\ [1ex]
 Primary $q_1$\dotfill            &  $0.325^{+0.064}_{-0.064}$             & $G(0.198, 0.1)$ \\ [1ex]
 Secondary $q_1$\dotfill          &  $0.554^{+0.104}_{-0.090}$             & $G(0.344, 0.1)$ \\ [1ex]
 $A_1$\dotfill                    &  $0.66^{+0.24}_{-0.34}$                & $G(0.5, 0.3)$ \\ [1ex]
 $A_2$\dotfill                    &  $0.35^{+0.12}_{-0.12}$                & $G(0.5, 0.3)$ \\ [1ex]
 $\ell_3$\dotfill                 &  $0.000^{+0.025}_{-0.000}$             & [$-$10, 0]* \\ [1ex]
 $\gamma$ (\kms)\dotfill          &  $+41.338^{+0.057}_{-0.038}$           & [30, 50] \\ [1ex]
 $K_1$ (\kms)\dotfill             &  $58.964^{+0.058}_{-0.045}$            & [1, 180] \\ [1ex]
 $K_2$ (\kms)\dotfill             &  $137.2^{+2.5}_{-2.1}$                 & [1, 180] \\ [1ex]
 $\Delta{\rm RV}$ (\kms)\dotfill  &  $-0.8^{+1.5}_{-1.5}$                  & [$-$10, 10] \\ [1ex]
 $P_{\rm out}$ (days)\dotfill     &  $463.5^{+2.4}_{-2.5}$                 & [200, 800] \\ [1ex]
 $T_{\rm peri}$ (HJD$-$2,400,000)\dotfill  &  $58076.9^{+5.4}_{-4.2}$      & [57900, 58600] \\ [1ex]
 $K_{\rm out}$ (\kms)\dotfill     &  $5.436^{+0.073}_{-0.074}$             & [1, 180] \\ [1ex]
 $\sqrt{e_{\rm out}} \cos\omega_{\rm out}$\dotfill  &  $-0.016^{+0.024}_{-0.033}$ & [$-$1, 1] \\ [1ex]
 $\sqrt{e_{\rm out}} \sin\omega_{\rm out}$\dotfill  &  $+0.437^{+0.019}_{-0.020}$ & [$-$1, 1] \\ [1ex]
 $f_{K2}$\dotfill                 &  $0.980^{+0.044}_{-0.044}$             & [$-$5, 1]* \\ [1ex]
 $f_1$\dotfill                    &  $1.06^{+0.16}_{-0.11}$                & [$-5$, 5]* \\ [1ex]
 $f_2$\dotfill                    &  $0.945^{+0.162}_{-0.092}$             & [$-5$, 5]* \\ [1ex]
\noalign{\hrule} \\ [-1.5ex]
\multicolumn{3}{c}{Derived quantities} \\ [1ex]
\noalign{\hrule} \\ [-1.5ex]
 $r_1$\dotfill                    &  $0.23386^{+0.00099}_{-0.00091}$       & \nodata \\ [1ex]
 $r_2$\dotfill                    &  $0.06087^{+0.00085}_{-0.00075}$       & \nodata \\ [1ex]
 $i$ (degree)\dotfill             &  $89.91^{+0.09}_{-1.02}$               & \nodata \\ [1ex]
 Eclipse duration (hour)\dotfill  &  $6.292^{+0.024}_{-0.021}$             & \nodata \\ [1ex]
 Primary $u_1$\dotfill            &  $0.265^{+0.044}_{-0.044}$             & \nodata \\ [1ex]
 Secondary $u_1$\dotfill          &  $0.58^{+0.10}_{-0.13}$                & \nodata \\ [1ex]
 $J_{\rm ave}$\dotfill            &  $0.096^{+0.013}_{-0.013}$             & \nodata \\ [1ex]
 $\ell_2/\ell_1$\dotfill          &  $0.00751^{+0.00066}_{-0.00065}$       & \nodata \\ [1ex]
 $e_{\rm out}$\dotfill            &  $0.191^{+0.018}_{-0.016}$             & \nodata \\ [1ex]
 $\omega_{\rm out}$ (degree)\dotfill  &  $92.2^{+4.3}_{-3.1}$              & \nodata 
\enddata

\tablecomments{The values listed correspond to the mode of the
  respective posterior distributions, and the uncertainties represent
  the 68.3\% credible intervals. Priors in square brackets are uniform
  over the specified ranges, except those for $\ell_3$, $f_{\ktwo}$,
  $f_1$, and $f_2$ (marked with asterisks), which are log-uniform. For
  the first order limb-darkening coefficients and the albedos the
  priors were Gaussian, indicated above as $G({\rm mean}, \sigma)$.}

\end{deluxetable}
\setlength{\tabcolsep}{6pt}

\begin{figure*}
\epsscale{1.15}
\includegraphics[width=0.515\textwidth,angle=270]{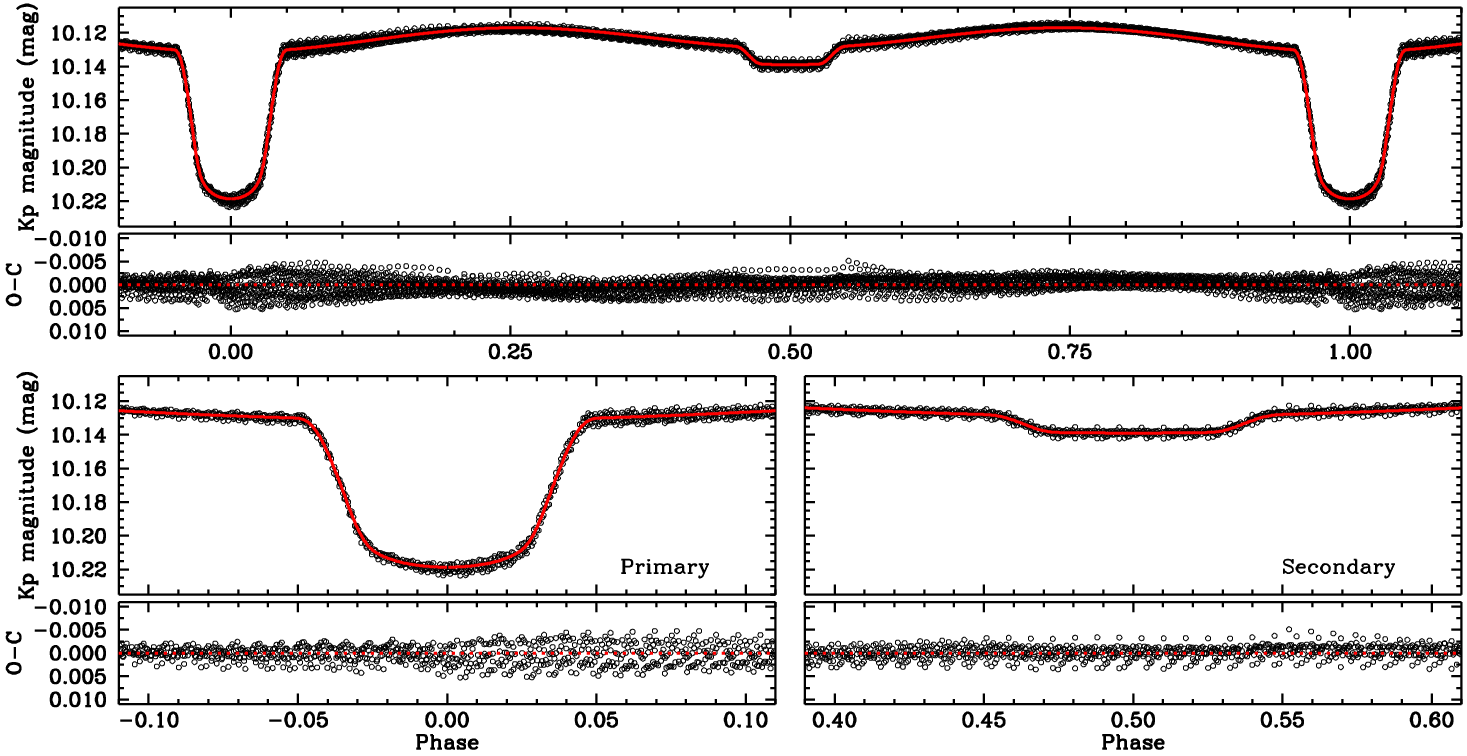}

\figcaption{\ktwo\ observations of \epic\ along with our adopted
  model. Enlargements of the eclipses are shown at the bottom.
  Residuals in magnitude units are displayed on an expanded scale
  below each panel.\label{fig:LC}}

\end{figure*}

The observations and final model are shown in Figure~\ref{fig:LC}. As
noted earlier, we attribute the considerable scatter of the residuals
(just under 1.6 mmag) to photometric modulation from spots rotating in
and out of view. The nature of this scatter is highly correlated
(``red'') noise, which raises at least two concerns. On the one hand,
it could introduce possibly significant biases in the results. On the
other, it will generally cause the formal uncertainties from our MCMC
analysis to be underestimated.  We now address each of these issues in
turn.

To gain an understanding of the extent to which these distortions may
affect the fitted parameters, we divided the complete \ktwo\ data set
into 29 separate cycles each containing one primary and one secondary
eclipse (for an average of 125 data points per cycle), with the last
cycle including an extra primary eclipse. We repeated the analysis
independently for each cycle in the same way as above, except that we
added a 4-term Fourier series to the model (9 extra parameters) in
order to at least partially account for the distortions, and we used
only the photometry for computational expediency.  The fundamental
period was kept fixed at the orbital period, which along with the mass
ratio was adopted from our model results in Table~\ref{tab:mcmc}.  The
median value for each parameter over the 29 data segments, and the
corresponding 68.3\% confidence intervals, are given in
Table~\ref{tab:cycles}.  Comparison with the values in
Table~\ref{tab:mcmc} indicates very good agreement for the geometric
parameters $r_1+r_2$, $k$, and $\cos i$, which are the most relevant
here. From this we conclude that any detrimental effect of the
lightcurve distortions seems to average out over the 29 cycles, at
least in this particular case. The agreement is in fact also good for
all other parameters in Table~\ref{tab:cycles} except for the limb
darkening coefficient of the secondary, and to a lesser degree its
albedo. These may well be biased in Table~\ref{tab:mcmc}, but have no
influence on any other results. The addition of the 4-term Fourier
series to the model for each cycle clearly improves the solutions
considerably, reducing the typical scatter by a factor of more than 6
compared to our original fit, from about 1.6~mmag to
$\sim$0.4~mmag. This is still larger than what the instrument is
capable of delivering because the lightcurve distortions are far from
regular.

\setlength{\tabcolsep}{5pt}
\begin{deluxetable}{lc}
\tablewidth{0pc}

\tablecaption{Results from our cycle-by-cycle MCMC Analysis for
  \epic\ with the addition of a 4-term Fourier series to the
  model\label{tab:cycles}}

\tablehead{ \colhead{~~~~~~~~~~~Parameter~~~~~~~~~~~} & \colhead{Value} }
\startdata
 $J$\dotfill                      &  $0.0907^{+0.0074}_{-0.0080}$      \\ [1ex]
 $r_1+r_2$\dotfill                &  $0.29474^{+0.00063}_{-0.00068}$   \\ [1ex]
 $k$\dotfill                      &  $0.25962^{+0.00092}_{-0.00067}$   \\ [1ex]
 $\cos i$\dotfill                 &  $0.00197^{+0.00053}_{-0.00093}$   \\ [1ex]
 Primary $q_1$\dotfill            &  $0.330^{+0.024}_{-0.015}$         \\ [1ex]
 Secondary $q_1$\dotfill          &  $0.3514^{+0.0101}_{-0.0085}$      \\ [1ex]
 $A_1$\dotfill                    &  $0.508^{+0.031}_{-0.019}$         \\ [1ex]
 $A_2$\dotfill                    &  $0.533^{+0.058}_{-0.067}$         \\ [1ex]
 $\ell_3$\dotfill                 &  $0.00015^{+0.00005}_{-0.00004}$   \\ [1ex]
 $f_{K2}$\dotfill                 &  $0.151^{+0.020}_{-0.026}$         \\ [1ex]
\noalign{\hrule} \\ [-1.5ex]
\multicolumn{2}{c}{Derived quantities} \\ [1ex]
\noalign{\hrule} \\ [-1.5ex]
 $r_1$\dotfill                    &  $0.23400^{+0.00059}_{-0.00053}$   \\ [1ex]
 $r_2$\dotfill                    &  $0.06066^{+0.00015}_{-0.00019}$   \\ [1ex]
 $i$ (degree)\dotfill             &  $89.887^{+0.053}_{-0.030}$        \\ [1ex]
 Primary $u_1$\dotfill            &  $0.270^{+0.022}_{-0.013}$         \\ [1ex]
 Secondary $u_1$\dotfill          &  $0.4204^{+0.0086}_{-0.0118}$      \\ [1ex]
 $J_{\rm ave}$\dotfill            &  $0.0917^{+0.0062}_{-0.0077}$      \\ [1ex]
 $\ell_2/\ell_1$\dotfill          &  $0.00735^{+0.00036}_{-0.00029}$    
\enddata

\tablecomments{The values listed correspond to the median of the
  results for the 29 independent orbital cycles. The uncertainties
  represent the 68.3\% credible intervals.}

\end{deluxetable}
\setlength{\tabcolsep}{6pt}

To address the possibility of underestimated uncertainties, we carried
out a residual permutation exercise in a similar way as done for our
previous two studies in Paper~I and Paper~II. We generated many
synthetic data sets by shifting the photometric residuals from our
adopted model by an arbitrary number of time indices, and adding them
back into the model curve at each time of observation (with
wrap-around). We then performed a new MCMC analysis on each set, in
each case using slightly perturbed values for quantities that had been
held fixed in our original analysis (the second-order limb darkening
coefficients, and the gravity darkening coefficients). The perturbed
quantities were generated by adding Gaussian noise to the values from
theory with standard deviations of 0.10 for $u_2$ and 0.05 for $y_1$
and $y_2$. We repeated this 100 times, and adopted the scatter
(standard deviation) of the resulting distribution for each fitted
parameter as a more realistic measure of the uncertainty. These
numbers were added quadratically to the internal errors from our
original MCMC analysis, resulting in the final uncertainties reported
in Table~\ref{tab:mcmc}. The parameters that had their internal errors
inflated the most are $J$, $k$, $q_1$ for the primary, and the albedo
$A_2$ for the secondary, by factors typically ranging from 3 to about
7. In other cases, the extra error is similar to or smaller than the
internal errors.

\section{Absolute dimensions}
\label{sec:dimensions}

The physical properties we infer for the components of \epic\ are
presented in Table~\ref{tab:dimensions}, in which the values listed
correspond to the mode of the posterior distributions calculated by
directly combining the chains of adjusted parameters in the top
section of Table~\ref{tab:mcmc}. The uncertainties represent the
68.3\% confidence intervals. The precision in the absolute masses is
4.2\% for the primary and 2.3\% for the secondary, while the radii
have errors of 1.2\% and 2.0\%, respectively.  Included among the
physical parameters are the luminosities, the absolute bolometric and
visual magnitudes, and the distance to the system (determined to about
7.7\%). For derived quantities involving external information, those
external quantities (effective temperatures, bolometric corrections,
interstellar reddening, and the apparent visual magnitude of the
system; see below) were assumed to be distributed normally and
independently for combining them with the chains of adjusted
parameters.

\begin{deluxetable}{lcc}
\tablewidth{0pc}
\tablecaption{Physical Properties of \epic \label{tab:dimensions}}
\tablehead{ \colhead{~~~~~~~~~~Parameter~~~~~~~~~~} & \colhead{Primary} & \colhead{Secondary} }
\startdata
 $M$ ($\mathcal{M}_{\sun}^{\rm N}$)\dotfill &  $1.509^{+0.063}_{-0.056}$     &  $0.649^{+0.015}_{-0.014}$ \\ [1ex]
 $R$ ($\mathcal{R}_{\sun}^{\rm N}$)\dotfill &  $2.505^{+0.026}_{-0.031}$     &  $0.652^{+0.013}_{-0.012}$ \\ [1ex]
 $\log g$ (dex)\dotfill                     &  $3.8216^{+0.0093}_{-0.0090}$  &  $4.624^{+0.011}_{-0.012}$ \\ [1ex]
 $q \equiv M_2/M_1$\dotfill                 &          \multicolumn{2}{c}{$0.4296^{+0.0070}_{-0.0077}$}   \\ [1ex]
 $a$ ($\mathcal{R}_{\sun}^{\rm N}$)\dotfill &          \multicolumn{2}{c}{$10.69^{+0.13}_{-0.12}$}        \\ [1ex]
 $T_{\rm eff}$ (K)\dotfill                  &  $6180 \pm 100$                &  $4010 \pm 170$            \\ [1ex]
 $L$ ($L_{\sun}$)\dotfill                   &  $8.19^{+0.61}_{-0.52}$        &  $0.096^{+0.021}_{-0.013}$ \\ [1ex]
 $M_{\rm bol}$ (mag)\dotfill                &  $2.442^{+0.077}_{-0.072}$     &  $7.22^{+0.21}_{-0.17}$    \\ [1ex]
 $BC_V$ (mag)\dotfill                       &  $-0.023 \pm 0.100$            &  $-1.131 \pm 0.100$        \\ [1ex]
 $M_V$ (mag)\dotfill                        &  $2.47^{+0.12}_{-0.12}$        &  $8.37^{+0.22}_{-0.21}$    \\ [1ex]
 $v_{\rm sync} \sin i$ (\kms)\tablenotemark{a}\dotfill  &  $46.01^{+0.48}_{-0.56}$  &  $11.98^{+0.23}_{-0.23}$ \\ [1ex]
 $v \sin i$ (\kms)\tablenotemark{b}\dotfill &  $49 \pm 3$                    &  12 (adopted)              \\ [1ex]
 $E(B-V)$ (mag)\dotfill                     &          \multicolumn{2}{c}{0.119~$\pm$~0.025}              \\ [1ex]
 $A_V$ (mag)\dotfill                        &          \multicolumn{2}{c}{0.369~$\pm$~0.078}              \\ [1ex]
 Dist.\ modulus (mag)\dotfill               &          \multicolumn{2}{c}{$7.33^{+0.15}_{-0.15}$}         \\ [1ex]
 Distance (pc)\dotfill                      &          \multicolumn{2}{c}{$291^{+22}_{-18}$}              \\ [1ex]
 $\pi$ (mas)\dotfill                        &          \multicolumn{2}{c}{$3.39^{+0.26}_{-0.21}$}         \\ [1ex]
 $\pi_{Gaia/{\rm DR2}}$ (mas)\tablenotemark{c}\dotfill  &          \multicolumn{2}{c}{$3.671 \pm 0.052$}
\enddata

\tablecomments{The masses, radii, and semimajor axis $a$ are expressed
  in units of the nominal solar mass and radius
  ($\mathcal{M}_{\sun}^{\rm N}$, $\mathcal{R}_{\sun}^{\rm N}$) as
  recommended by 2015 IAU Resolution B3 \citep[see][]{Prsa:2016}, and
  the adopted solar temperature is 5772~K (2015 IAU Resolution
  B2). Bolometric corrections are from the work of \cite{Flower:1996},
  with conservative uncertainties of 0.1~mag, and the bolometric
  magnitude adopted for the Sun appropriate for this $BC_V$ scale is
  $M_{\rm bol}^{\sun} = 4.732$ \citep[see][]{Torres:2010}. See text
  for the source of the reddening. For the apparent visual magnitude
  of \epic\ out of eclipse we used $V = 10.164 \pm 0.026$
  \citep{Henden:2015}. The flux of the tertiary component is ignored
  here.}

\tablenotetext{a}{Synchronous projected rotational velocity assuming
  spin-orbit alignment.}

\tablenotetext{b}{Measured projected rotational velocity for the
  primary.}

\tablenotetext{c}{A global parallax zero-point correction of
  $+0.029$~mas has been added to the parallax \citep{Lindegren:2018a},
  and 0.021~mas added in quadrature to the internal error
  \citep[see][]{Lindegren:2018b}.}

\end{deluxetable}

In Section~\ref{sec:spectroscopy} we derived an estimate of the
primary temperature directly from our spectra ($6180 \pm 100$~K), but
had to use an adopted value for the much fainter secondary (4000~K)
for the radial-velocity determinations. Our analysis of the light
curve now provides an accurate way to measure the temperature ratio
(or difference, $\Delta T_{\rm eff}$) between the components, through
the central surface parameter $J$ (or the disk-integrated value
$J_{\rm ave}$; Table~\ref{tab:mcmc}). This, then, allows the secondary
temperature to be inferred. Such estimates are often quite accurate
because $J$ is closely related to the difference in depth between the
eclipses, which can be measured accurately. We obtain $\Delta T_{\rm
  eff} = 2170 \pm 140$~K, with which the secondary temperature becomes
$4010 \pm 170$~K. This is essentially the same as the value adopted in
Section~\ref{sec:spectroscopy} for the radial-velocity determinations,
justifying that choice a posteriori. We adopt this as the final
temperature of the secondary, and list it in
Table~\ref{tab:dimensions}.

The component temperatures may be used to derive an estimate of the
reddening, as done in our earlier studies of Paper~I and Paper~II. We
refer the reader to those sources for the details. Briefly, we
gathered standard photometry for the combined light of \epic\ in the
Tycho-2, Johnson, Sloan, 2MASS, and {\it Gaia\/} systems
\citep{Hog:2000, Henden:2015, Skrutskie:2006, Gaia:2018} and we
constructed 14 non-independent color indices. We then used
color/temperature calibrations by \cite{Casagrande:2010},
\cite{Huang:2015}, and \cite{Stassun:2019} to derive an average
photometric temperature for a range of reddening values, adjusting all
color indices appropriately at each value of $E(B-V)$.  Our adopted
reddening estimate is the one that provides a match to the
luminosity-weighted average temperature of the binary system. We
ignored the presence of the tertiary, as it is faint enough that it
will not affect the results. We obtained $E(B-V) = 0.119 \pm
0.025$~mag, corresponding to $A_V = 0.369 \pm 0.078$~mag for a ratio
of total to selective extinction of $R_V = 3.1$. Very similar values
were found for the other two eclipsing binaries studied previously in
the cluster.

\begin{figure}
\epsscale{1.15}
\plotone{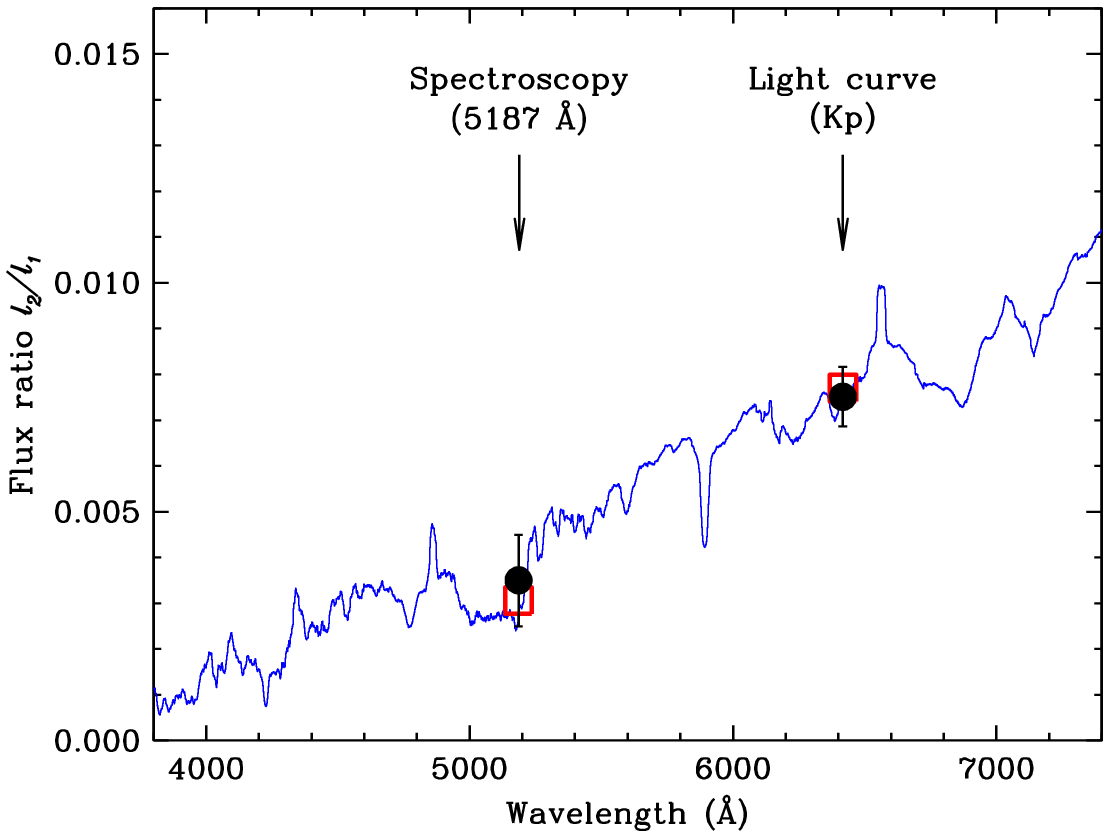}

\figcaption{Comparison of the predicted flux ratio between the
  components of \epic\ and our $\ell_2/\ell_1$ values measured
  spectroscopically and from the light curve analysis. Squares
  represent the calculated flux ratio integrated over the
  corresponding spectroscopic and \ktwo\ bandpasses, and the
  measurements are shown as circles with error bars. The calculated
  curve is based on model spectra by \cite{Husser:2013} for solar
  metallicity, normalized using our measured radius ratio $k =
  0.260$. For the primary we used $T_{\rm eff} = 6200$~K and $\log g =
  4.0$, and for the secondary $T_{\rm eff} = 4000$~K and $\log g =
  4.5$.\label{fig:fluxratio}}

\end{figure}

A consistency check on our effective temperatures and radii may be
obtained by comparing our measured flux ratios with predicted values
from synthetic spectra.  Figure~\ref{fig:fluxratio} shows the
predictions as a function of wavelength, using spectra by
\cite{Husser:2013} based on PHOENIX model atmospheres for temperatures
of 6200~K and 4000~K, near our best estimates for the binary
components. The normalization of the ratio of the spectra was carried
out with the radius ratio derived from our light curve analysis, $k =
0.260$. Our measured flux ratios from spectroscopy and from the
\ktwo\ photometry show very good agreement with the expected values,
supporting the accuracy of our determinations for \epic.

The measured projected rotational velocity for the primary star, $v
\sin i = 49 \pm 3$~\kms, agrees with the predicted value $v_{\rm
  sync}\sin i$ listed in Table~\ref{tab:dimensions}, which assumes
synchronous rotation and spin-orbit alignment. Given the $\sim$2.7~Gyr
age of the parent cluster, this is as expected from tidal theory
\citep[see, e.g.][]{Hilditch:2001}, which indicates synchronization
should occur on a timescale of less than 2~Myr. The negligible
eccentricity we find for the inner orbit is also consistent with the
expectation that tidal forces should circularize it on a timescale of
about 300~Myr, much shorter than the cluster age.

The elements we derive for the outer orbit, along with the absolute
masses for the binary components, imply a minimum mass for the
tertiary star of about 0.36~$M_{\sun}$. If it is a main sequence star,
the lack of detection in our spectra implies a mass that can be no
larger than that of the secondary, or $\sim$0.65~$M_{\sun}$.  This, in
turn, gives a lower limit for the inclination angle of the outer orbit
of about 34\arcdeg. Alternatively, the tertiary may be a white dwarf.

\subsection{Rotation and activity}
\label{sec:rotation}

\begin{figure}
\epsscale{1.15}
\plotone{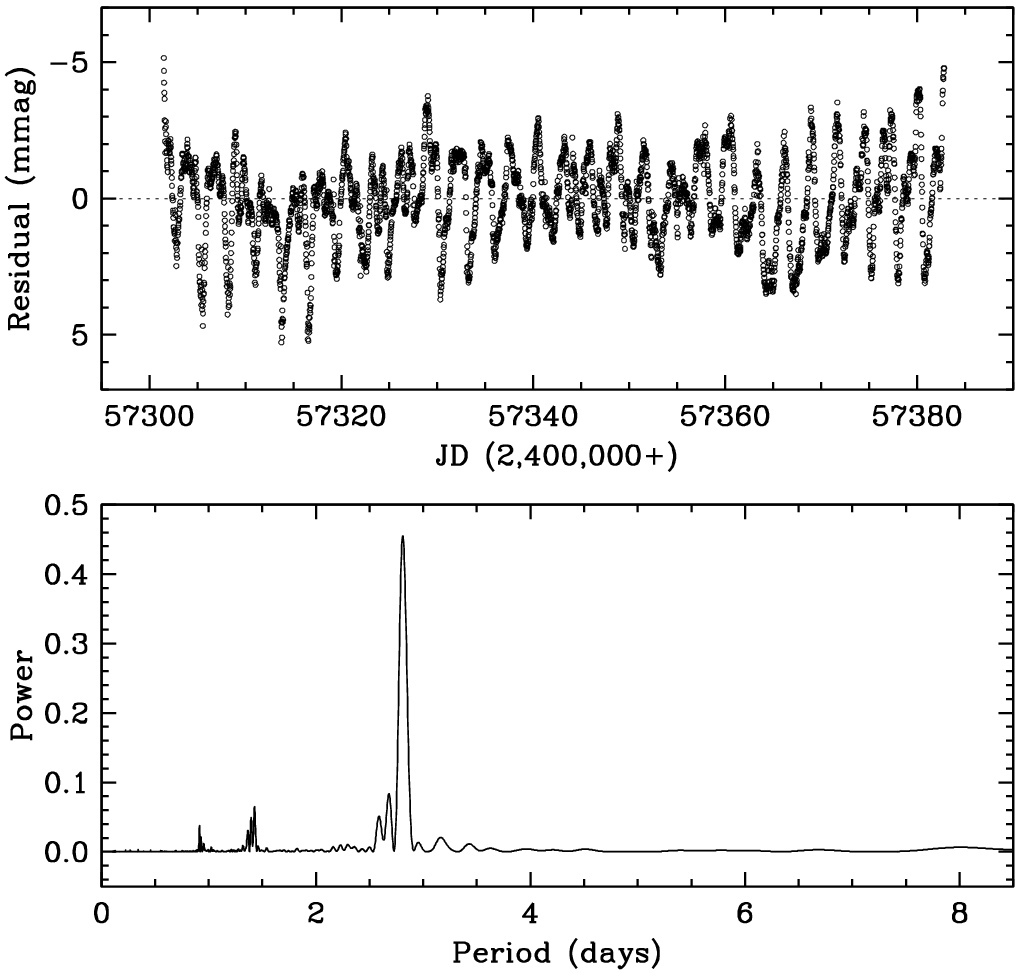}

\figcaption{Residuals from our best fit to the \ktwo\ photometry of
  \epic\ showing modulations presumably due to spots. The
  corresponding periodogram shown at the bottom features a dominant
  peak at a period of $P_{\rm rot} = 2.81 \pm
  0.04$~days. \label{fig:periodogram}}

\end{figure}

Oscillations in the \ktwo\ photometry of \epic\ are obvious in the
residuals from our light curve analysis, and are seen as a function of
time in the top panel of Figure~\ref{fig:periodogram}. The
oscillations are rather irregular (see also Figure~\ref{fig:spots}),
and display a peak-to-peak amplitude close to 10 mmag.  This amplitude
is in line with those seen in other late F dwarfs, as reported by
\cite{Giles:2017}.  A Lomb-Scargle periodogram of the residuals shows
a main peak at a period of $2.81 \pm 0.04$~days (bottom panel of
Figure~\ref{fig:periodogram}), which we interpret as a rotational
signature caused by one or more spots or spot regions. The uncertainty
was estimated from the half width of the peak at half maximum. This
period is marginally longer than the 2.75-day orbital period of the
binary, which could imply subsynchronous rotation, or may also be a
consequence of solar-like differential rotation, with spots located at
intermediate or high latitudes rotating more slowly than the
equator. In view of the spectroscopic evidence ($v \sin i$ of the
primary) presented earlier for spin-orbit synchronization, we are
inclined to favor the latter interpretation. The spots are most likely
located on the primary star, as assuming that they are on the
secondary would imply a rather unusual intrinsic amplitude exceeding
one magnitude, because of the large brightness dilution factor
($\ell_2/\ell_1 \approx 0.0075$).

While spot modulation is often associated with other indicators of
stellar activity, examination of the spectra for \epic\ has revealed
no evidence of \ion{Ca}{2} H and K emission, or variable H$\alpha$
equivalent widths. This is perhaps consistent with the relatively
small amplitude of the photometric variations. We note also that, as
far as we can tell, the object does not appear to have been detected
as an X-ray source \citep[e.g., by {\it ROSAT};][]{Voges:1999} or as a
source of ultraviolet radiation \citep[{\it
    GALEX};][]{Bianchi:2011}. It would not be surprising if the faint
secondary star were somewhat active as well, despite the old age of
the system, given its moderately large expected rotational velocity of
about 12~\kms\ (Table~\ref{tab:mcmc}).

\section{Comparison with theory}
\label{sec:models}

\begin{figure}
\epsscale{1.15}
\plotone{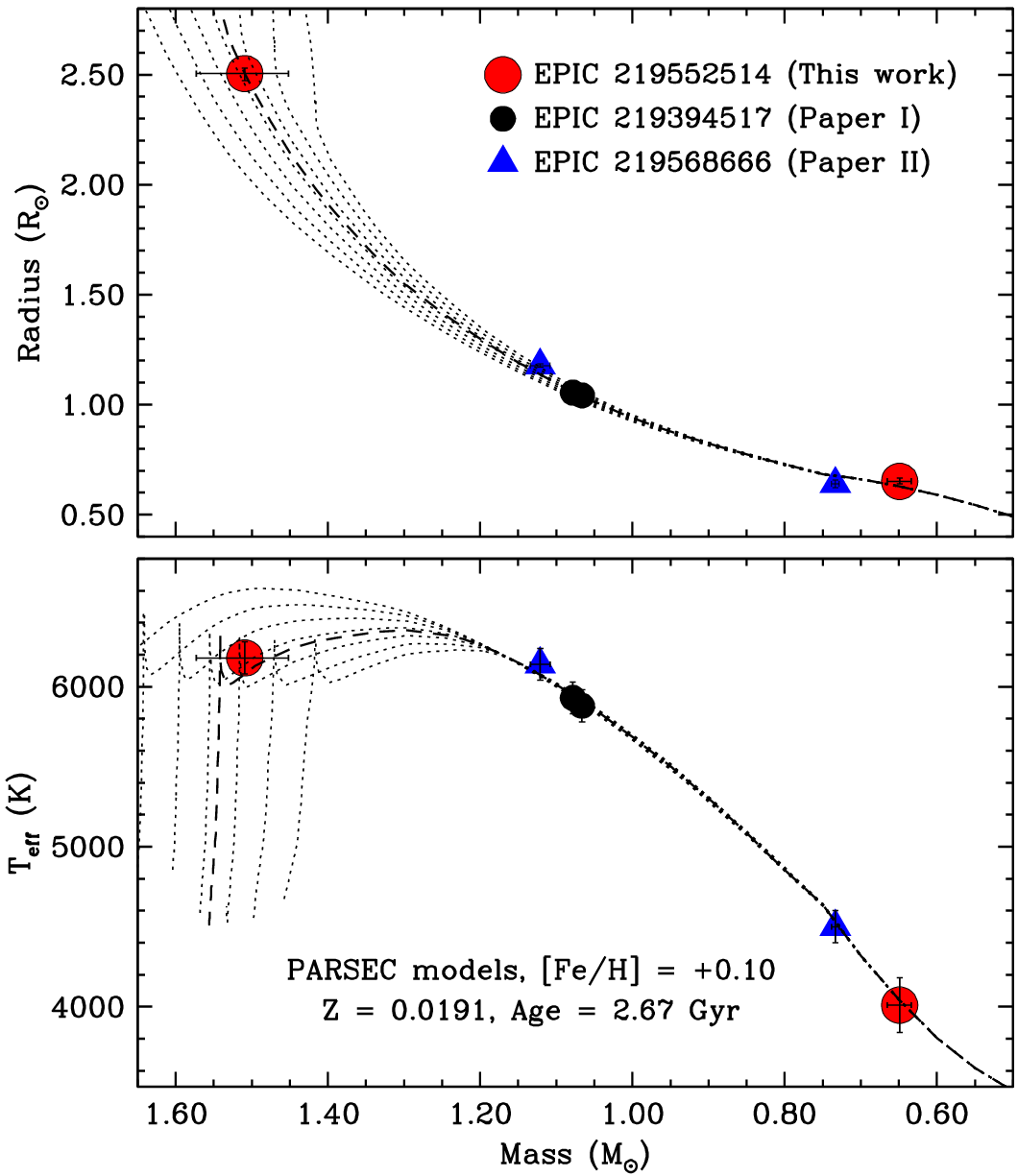}

\figcaption{Physical properties of \epic\ compared against model
  isochrones from the PARSEC v1.2S series \citep{Chen:2014} for the
  metallicity of the cluster, ${\rm [Fe/H]} = +0.10$. The dotted lines
  in both panels represent isochrones for ages between 2.0 and 3.2~Gyr
  in steps of 0.2~Gyr, and the heavy dashed line corresponds to an age
  of 2.67~Gyr that fits the measured masses and radii best. Results
  for the two eclipsing binaries studied previously in Paper~I and
  Paper~II are shown as well. \label{fig:parsec}}

\end{figure}

The masses, radii, and temperatures of the binary components of
\epic\ are compared in Figure~\ref{fig:parsec} against models of
stellar evolution from the PARSEC v1.2S series by \cite{Chen:2014}.
Isochrones in both the mass-radius and mass-temperature diagrams are
shown for the age range 2.0--3.2~Gyr in steps of 0.2~Gyr, with the
heavy dashed line representing the best fit. The corresponding age is
$2.67^{+0.39}_{-0.55}$~Gyr, in which the uncertainty is dominated by
the error in the primary mass. Uncertainties in the adopted chemical
composition of \rup\ \citep[${\rm [Fe/H]} = +0.10 \pm 0.04$;
  see][]{Curtis:2018} contribute an additional 0.13~Gyr to the error
budget for the age.

Because the age determination for \epic\ is only sensitive to the
radius of the primary star (the secondary evolves too slowly), the
best-fit isochrone matches $R_1$ precisely. The secondary radius
appears slightly larger than predicted (by 3.7\%), although the
deviation is less than twice its uncertainty and may not be
significant. On the other hand, the effective temperatures of both
components are consistent with predictions from theory. Many cool main
sequence stars such as the secondary have shown discrepancies with
standard stellar evolution models that are believed to be caused by
stellar activity \citep[see, e.g.,][]{Torres:2013}. They tend to be
larger and cooler than predicted.  However, in this case a reasonably
good agreement between the [$R$, $T_{\rm eff}$] measurements for the
secondary and these particular models is expected a priori because the
PARSEC v1.2S models have been adjusted by changing the
temperature-opacity relation in such a way as to match the average
measured properties of low-mass stars \citep[see][]{Chen:2014}.

The above age determination for \epic\ agrees well with our estimates
for the eclipsing binaries EPIC~219394517 and EPIC~219568666 from our
earlier studies (Paper~I and Paper~II), whose physical properties are
also shown in the figure.\footnote{As discussed in Paper~II, the
  individual component radii for EPIC~219568666 were reported to be
  slightly affected by systematic errors in the radius ratio $k$,
  although the age could still be determined accurately using instead
  the sum of the radii, which is unaffected.} The ages we reported for
those two objects using the same PARSEC models as above are $2.65 \pm
0.25$ and $2.76 \pm 0.61$~Gyr. Other models with different physical
ingredients lead to slightly different ages. For example, using the
MIST models of \cite{Choi:2016}, we obtain a marginally younger age
for \epic\ of 2.51~Gyr.

\begin{figure}
\epsscale{1.15}
\plotone{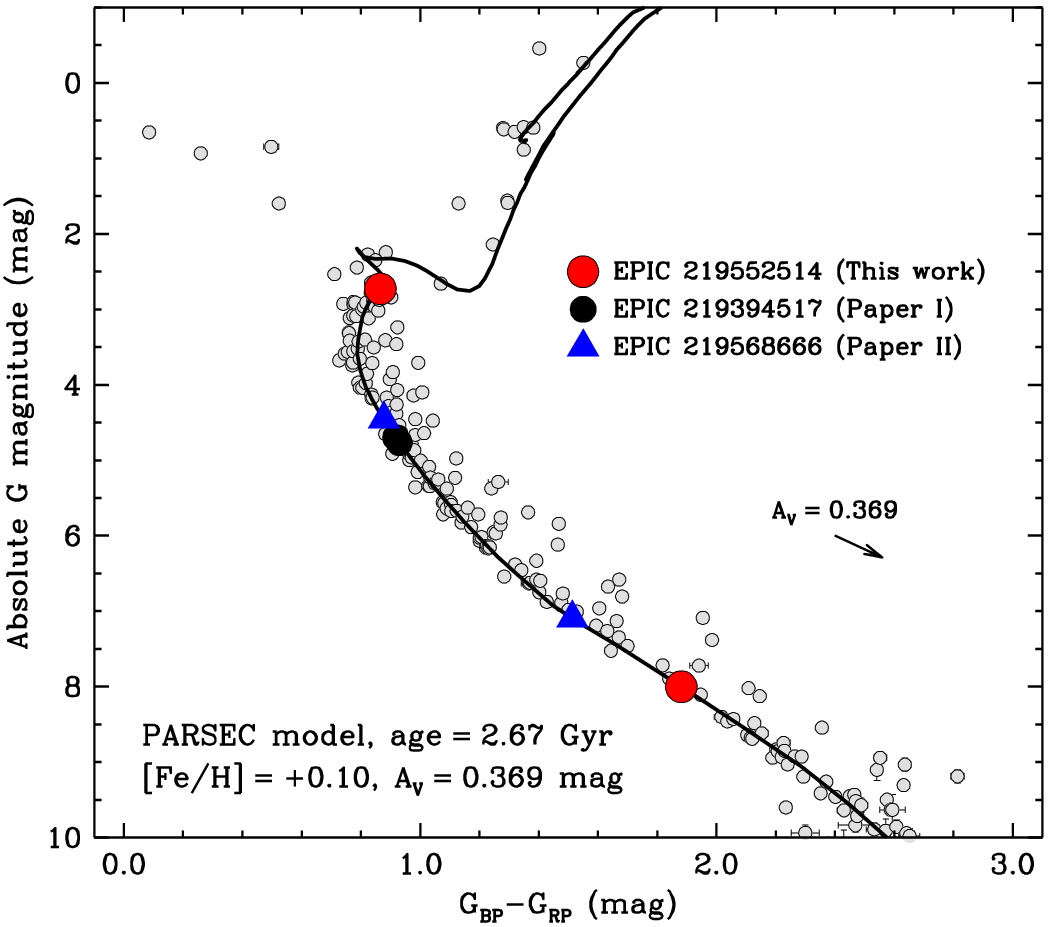}

\figcaption{Color-magnitude diagram for \rup\ based on the
  measured $G$ magnitudes, $G_{\rm BP}-G_{\rm RP}$ colors, and
  parallaxes from the {\it Gaia}/DR2 catalog. Also shown is a model
  isochrone from the PARSEC series \citep{Chen:2014} for the
  metallicity of the cluster and the age that best fits the properties
  of \epic\ (Figure~\ref{fig:parsec}).  Reddening and extinction
  corrections have been applied to the model (reddening vector
  indicated with an arrow). The same symbols as in
  Figure~\ref{fig:parsec} are used to mark the locations on the main
  sequence of the stars in \epic\ and the other two eclipsing binaries
  in the cluster studied previously. \label{fig:cmd}}

\end{figure}

The evolved status of the primary star places it at the very end of
the main sequence for the cluster. This may be seen in the
color-magnitude diagram of Figure~\ref{fig:cmd}, which shows other
member stars from the {\it Gaia}/DR2 catalog along with the 2.67~Gyr
PARSEC isochrone corrected for reddening and extinction. The absolute
magnitudes for the other members were calculated using their
individual parallaxes. The two previously studied binaries in
\rup\ are marked on the isochrone as well, at the locations
expected from their measured masses.

\section{Final remarks}
\label{sec:discussion}

\epic\ is special for being located near the turnoff of \rup,
making it the system most sensitive to age among the known eclipsing
binaries in the cluster. Our accurate mass and radius determinations
have allowed an accurate age to be inferred for the binary
($2.67^{+0.39}_{-0.55}$~Gyr) based on the PARSEC models. This is in
excellent agreement with estimates for the two previously studied
eclipsing systems in the cluster that used the same models. The
weighted average of the three determinations is $2.67 \pm
0.21$~Gyr. The precision of our current age for \epic\ is limited by
the uncertainty in the mass of the primary star (4.2\%), which in turn
is caused mostly by the reduced precision of the radial velocities of
the secondary star on account of its faintness ($\sigma_{\rm RV}
\sim$8.5~\kms, on average). Additional spectra with higher
signal-to-noise ratios would help to reduce these statistical errors.

Our distance estimate for \epic\ is similar to those inferred in
Paper~I and Paper~II, and implies a parallax
($3.39^{+0.26}_{-0.21}$~mas) that is marginally lower than the one
reported in the {\it Gaia}/DR2 catalog ($3.671 \pm 0.051$~mas),
although still consistent within the errors. It is possible that part
of the difference is due to the fact that the system is found here to
be triple, whereas {\it Gaia\/} has so far been treating the object as
a single star. The discovery that \epic\ is a triple system is in fact
not surprising, as it has been found that the vast majority of
spectroscopic binaries with periods under 3 days have additional
companions \citep[96\%, according to][]{Tokovinin:2006}.

The nature of the third component, i.e., whether it is a main-sequence
star or a white dwarf, is undetermined from the present data. In
either case, this distant companion may play a role in the dynamical
evolution of the system, modulating the eccentricity of the inner
eclipsing binary (which we currently find to be consistent with having
a circular orbit) as well as modulating the relative inclination angle
between the inner and outer orbital planes \cite[Kozai-Lidov
  oscillations; see, e.g.,][]{Naoz:2016}. The latter can potentially
change the eclipse depths, or temporarily cause them to cease
altogether. This would be expected to occur on timescales that are
much longer than the orbital periods. We find the system to be
dynamically stable according to the criteria of \cite{Eggleton:1995}
and \cite{Mardling:2001}, for any reasonable mass of the third star
and any relative inclination of the orbital planes.

If the tertiary is a white dwarf, constraints on its mass and that of
its progenitor may be obtained from the same PARSEC isochrone used
earlier for the age and metallicity of \rup. We find a lower limit for
the progenitor mass of 1.59~$M_{\sun}$, and a corresponding lower
limit for the present-day white dwarf mass of 0.60~$M_{\sun}$. This
would imply significant mass loss from the third component, which may
have left some observable trace on the system. One possiblity would be
chemical polution of the eclipsing binary components. This could be
pursued through a detailed spectroscopic analysis of the primary star.

We estimate the angular size of the outer orbit to be about 5.7~mas at
the distance of \rup, which is inside the reach of our NRM
observations described in Section~\ref{sec:imaging}.  While {\it Gaia}
cannot spatially resolve the tertiary, in principle it should be
capable of detecting the motion of the eclipsing binary in the outer
orbit with a period of 463 days. We estimate this motion should have a
semimajor axis between 0.8 and 1.3~mas, depending on the outer
inclination angle, assuming the tertiary contributes negligibly to the
total light. Disappointingly, the {\it Gaia}/DR2 data for this system
show no sign of excess astrometric noise (which could otherwise be an
indication of unmodeled motion), although it is still possible that
the 463 day signal may emerge or be recoverable by the end of the
mission, particularly with the knowledge we now have. In that case,
{\it Gaia\/} should be able to measure the inclination angle of the
outer orbit. When combined with the elements of our spectroscopic
orbit, this angle would then immediately allow a determination of the
dynamical mass of the tertiary component.


\begin{acknowledgements}

The spectroscopic observations of \epic\ were gathered with the help
of P.\ Berlind, M.\ Calkins, and G.\ Esquerdo. J.\ Mink is thanked for
maintaining the CfA echelle database. The anonymous referee provided
helpful comments on the original manuscript.  G.T.\ acknowledges
partial support from NASA's Astrophysics Data Analysis Program through
grant 80NSSC18K0413, and to the National Science Foundation (NSF)
through grant AST-1509375. J.L.C.\ is supported by the NSF Astronomy
and Astrophysics Postdoctoral Fellowship under award AST-1602662, and
by NASA under grant NNX16AE64G issued through the \ktwo\ Guest
Observer Program (GO~7035).
The research has
made use of the SIMBAD and VizieR databases, operated at the CDS,
Strasbourg, France, and of NASA's Astrophysics Data System Abstract
Service.
The work has also made use of data from the European Space Agency
(ESA) mission {\it Gaia} (\url{https://www.cosmos.esa.int/gaia}),
processed by the {\it Gaia} Data Processing and Analysis Consortium
(DPAC,
\url{https://www.cosmos.esa.int/web/gaia/dpac/consortium}). Funding
for the DPAC has been provided by national institutions, in particular
the institutions participating in the {\it Gaia} Multilateral
Agreement. The computational resources used for this research include
the Smithsonian Institution's ``Hydra'' High Performance Cluster.

\end{acknowledgements}

\end{document}